\documentclass[12pt, a4paper]{article}

\usepackage{amsmath}
\usepackage{array}
\usepackage{amsfonts}
\usepackage{amssymb}
\usepackage{graphicx, rotating}
\usepackage{epsfig}
\usepackage{latexsym}
\usepackage{graphicx}
\usepackage{color}
\usepackage{amsmath,bm,amssymb}
\usepackage{units}
\usepackage{cite}
\usepackage{slashed}
\usepackage{hyperref}
\usepackage{array}
\usepackage{booktabs}
\usepackage{authblk} 

\newcommand{\lp}{\left(}
\newcommand{\rp}{\right)}
\newcommand{\nn}{\nonumber}
\newcommand{\be}{\begin{equation}}
\newcommand{\ee}{\end{equation}}
\newcommand{\bea}{\begin{eqnarray}}
\newcommand{\eea}{\end{eqnarray}}

\begin{document}

\begin{titlepage}

\title{\vspace{-2.5cm}
\begin{flushright}
\normalsize DESY 20-174\\
TUM-HEP-1294/20\\
TTK-20-36 
\end{flushright}
\vspace{0.5cm}
\Large \bf Neutrino mass bounds from confronting an effective model with BOSS Lyman-$\alpha$ data}

\author[1]{\large Mathias~Garny \thanks{\href{mailto:mathias.garny@tum.de}{mathias.garny@tum.de}}}
\affil[1]{\normalsize \it Technical University Munich, Physics Department, James Franck Stra\ss e 1, 85748~Garching, Germany}

\author[2]{\large Thomas~Konstandin \thanks{\href{mailto:thomas.konstandin@desy.de}{thomas.konstandin@desy.de}}}
\affil[2]{\normalsize \it DESY, Notkestra\ss e 85, 22607 Hamburg, Germany}

\author[3,4]{\large Laura~Sagunski \thanks{\href{mailto:sagunski@itp.uni-frankfurt.de}{sagunski@itp.uni-frankfurt.de}}}
\affil[3]{\normalsize \it Institute for Theoretical Physics, Goethe University, 60438 Frankfurt am Main, Germany}
\affil[4]{\normalsize \it Institute for Theoretical Particle Physics and Cosmology~(TTK), RWTH~Aachen~University, 52056 Aachen, Germany}

\author[5,6,7,8]{\large Matteo~Viel \thanks{\href{mailto:viel@sissa.it}{viel@sissa.it}}}
\affil[5]{\normalsize \it SISSA, via Bonomea 265, 34136 Trieste, Italy }
\affil[6]{\normalsize \it IFPU, Institute for Fundamental Physics of the Universe, via Beirut 2, 34151 Trieste, Italy}
\affil[7]{\normalsize \it INFN, Sezione di Trieste, Via Valerio 2, I-34127 Trieste, Italy}
\affil[8]{\normalsize \it INAF - Osservatorio Astronomico di Trieste, Via G. B. Tiepolo 11, I-34143 Trieste, Italy}

\date{}

\maketitle

\begin{abstract}
We present an effective model for the one-dimensional Lyman-$\alpha$ flux power spectrum far above the baryonic Jeans scale.
The main new ingredient is constituted by a set of two parameters that encode the impact of small, highly non-linear scales on the one-dimensional power spectrum
on large scales, where it is measured by BOSS. We show that, by marginalizing over the model parameters that capture the impact of the
intergalactic medium, the flux power spectrum from both simulations and observations can be described with high precision.
The model displays a degeneracy between the neutrino masses and the (unknown, in our formalism) normalization of the flux power spectrum. 
This degeneracy can be lifted by calibrating one of the model parameters with simulation data, and using input from Planck CMB data. We demonstrate that this approach can be used to
extract bounds on the sum of neutrino masses with comparably low numerical effort, while allowing for a conservative treatment of uncertainties from the dynamics of the intergalactic medium. An explorative analysis yields an upper bound of $0.16\,$eV at $95\%$ C.L. when applied to BOSS data at $3\leq z\leq 4.2$. We also forecast that if the systematic and statistical errors will be reduced by a factor two the upper bound will become $0.1\,$eV at $95\%$ C.L., and $0.056\,$eV when assuming a $1\%$ error.
\end{abstract}

\bigskip

\end{titlepage}

\section{Introduction \label{sec:intro}} 

The Lyman-$\alpha$ forest is an eminent probe of cosmology since it tests the smallest scales in the matter power spectrum that are currently accessible to experiments. The Lyman-$\alpha$ forest contains the absorption lines of neutral hydrogen in a background emitted by high-redshift quasars. The absorption lines result from the intergalactic medium (IGM) along the line-of-sight between the quasars and the observer. Using this setup, it is possible to scrutinize fluctuations in the matter distribution below the Mpc scales, up to scales of hundreds of Mpc. This opens the possibility to probe cosmological models featuring a modified growth of structure on these scales, such as warm or mixed cold/hot dark matter. The latter scenario is realized within the standard cosmological model via the cosmic neutrino background, whose contribution to the total energy density is related to the sum of neutrino masses.
There are two main observables that have been used to perform cosmological investigations. On one side we can rely on the one-dimensional flux power spectrum, as extracted from a set of low, medium and high-resolution quasar spectra, to constrain cosmological parameters, warm dark matter scenarios, primordial black holes, fuzzy dark matter, dark matter-dark radiation interactions and neutrino masses; on the other hand the three-dimensional information, probing the largest scales, has been instrumental to provide tight measurements of distances at $z\sim 2.3$ with baryonic acoustic oscillations as seen in the auto-correlation function of the transmitted flux and the cross-correlation between  quasars and flux  as recently measured by the BOSS collaboration \cite{dumas20}.
This complementary view on dynamical growth and geometry of our Universe is very constraining for a large set of beyond the standard model scenarios of structure formation based on a cosmological constant and cold dark matter and is particularly important when these data sets are combined with other large scale structure tracers that probe larger scales and smaller redshifts.

The latest BOSS Lyman-$\alpha$ forest data \cite{Chabanier:2018rga} indicates some mild tension with Planck temperature and polarization measurements~\cite{Aghanim:2018eyx} within $\Lambda$CDM, that can be relaxed when allowing for a running spectral index. Furthermore, it has been demonstrated \cite{Palanque-Delabrouille:2019iyz} that strong bounds on the sum of neutrino masses can be derived by combining BOSS and Planck data even when including running ($\sum m_\nu \leq 0.13\,$eV at $95\%$C.L., compared to $0.10\,$eV without running). When using BOSS Lyman-$\alpha$ data only, also very large neutrino masses ($0.71\,$eV) are compatible with observations, which can be traced back to an approximate degeneracy with the amplitude $A_s$ of the primordial power spectrum (this is nicely demonstrated in \cite{Pedersen:2019ieb}).

Extracting information on cosmological models from Lyman-$\alpha$ forest observations requires a description of the IGM.
The distribution of the IGM depends on the initial distribution of dark matter in the Universe. Still, the dynamics of the IGM is very complex and non-linear such that extensive hydrodynamic simulations of the IGM are indispensable to arrive at a prediction of the flux power spectrum measured from the absorption spectra of a set of quasars \cite{Bird:2011rb,Pontzen:2014ena,bolton17,Rogers:2017bmq,Pedersen:2019ieb,Villasenor:2020huv}. However, these simulations are very expensive which poses a major challenge for parameter estimation and marginalization over the (a priori often unknown) parameters of the IGM dynamics. 
Furthermore, additional uncertainties can enter through an inhomogeneous UV background and a large mean free path length of UV photons in the IGM (see e.g.~\cite{Pontzen:2014ena,Cabass:2018hum}).

One possible strategy is to use a grid of simulations. This approach was followed in \cite{Palanque-Delabrouille:2014jca,Palanque-Delabrouille:2015pga,Palanque-Delabrouille:2019iyz} to extract bounds on neutrino masses and similarly bounds on warm dark matter as well as dark radiation models have been obtained \cite{irsic17,irsic17b,rogers20,murgia19,Murgia:2018now,Archidiacono:2019wdp}. Besides, bounds on ultra-light axions have been derived along the same lines \cite{Rogers:2020cup} using an Emulator based on simulation results. The impact of IGM uncertainties has also been emphasized in the analysis performed in the context of warm dark matter in \cite{Garzilli:2019qki}. 
All these approaches are ultimately based on a suite of high-resolution hydrodynamic simulations that incorporate all the relevant physical ingredients (e.g. \cite{bolton17}). These simulations allow to cover the variations of the one-dimensional flux power in terms of cosmological and astrophysical parameters and build a likelihood function and emulators that  ultimately allow to address the agreement with the data.

Alternatively, effective models can be used to extract information from the data. If this strategy can be demonstrated to be accurate at the required precision, it potentially allows one to evaluate a likelihood for many sets of parameters, including both cosmological as well as nuisance parameters that describe the uncertainties in the IGM dynamics and thermal state. Notice that such simple, semi-analytical, models have been found to be promising even in reproducing non-linear quantities like the flux probability distribution function \cite{seljak12} or in reaching small scales with a halo-model based approach \cite{irsicmcquinn18}, once they are calibrated with hydrodynamic simulations.

However, even though simple analytical approaches for the Lyman-$\alpha$ flux power spectrum exist \cite{Gnedin:1997td}, typically they fail to capture the dynamics of the IGM at the level of accuracy that is reached nowadays. In the current work, we follow in broad strokes the philosophy of analytical approaches, but deviate in two important points: First, we do not try to predict the different bias parameters of these models but obtain these parameters through calibration to either experimental or simulation data \cite{seljak12,cieplak16,hirata18,Arinyo-i-Prats:2015vqa}. Second, since the flux power spectrum is sensitive to very small scales, we follow an effective theory approach and parametrize our lack of knowledge by two ultra-violet (UV) parameters that carry this information (see section \ref{sec:review}). By construction, this approach leads to rather conservative bounds. A related approach along these lines has recently been demonstrated to yield meaningful constraints in the context of a full-shape analysis of BOSS galaxy clustering data \cite{Ivanov:2019hqk,Colas:2019ret}.

We build on the framework developed in \cite{Garny:2018byk}, where we showed that with six parameters, the model can describe BOSS data \cite{Palanque-Delabrouille:2013gaa} on the Lyman-$\alpha$ forest. In principle, data for even smaller scales exist, for example from XQ-100 \cite{Yeche:2017upn} and Hires/Mike~\cite{Viel:2013fqw}, but we focus on BOSS data which is on scales  much larger than the Jeans scale. This makes the analysis more robust, since smaller scales are more susceptible to the details of reionisation and the dynamics of the IGM. More importantly, the separation of scales is a prerequisite for the effective theory approach. 

The main aim of the present work is to demonstrate that an effective model  can provide an accurate description for the Lyman-$\alpha$ 1D flux power spectrum, and to assess its
predictive power when it comes to the sum of the neutrino masses. One roadblock in obtaining a competitive bound on neutrino masses are the degeneracies in the model. Therefore, we explore a hybrid approach, where one of the model parameters is calibrated by comparing to simulation results. As a first step, we marginalize over the remaining model parameters, describing the unknown IGM dynamics, for
a given $\Lambda$CDM model, with all cosmological parameters except for the neutrino mass kept fixed. We demonstrate that, within this framework, meaningful constraints on the neutrino mass can be derived, paving the way for an application in Monte Carlo parameter estimation techniques in the future. 

In section~\ref{sec:model}, we review the effective model for the 1D Lyman-$\alpha$ flux power spectrum and discuss how we incorporate massive neutrinos.
Next, we validate the model by comparing to simulation data in section~\ref{sec:val}, and introduce the hybrid approach. In section~\ref{sec:boss} we apply the framework to
the latest BOSS Lyman-$\alpha$ forest data \cite{Chabanier:2018rga}, and discuss the impact of various assumptions, before concluding in section~\ref{sec:sum}.

\section{Effective Lyman-$\alpha$ model for massive neutrinos\label{sec:model}} 

In this section we first review the effective model for the 1D flux power spectrum \cite{Garny:2018byk} on BOSS scales, and then discuss how we 
incorporate the effect of massive neutrinos.

\subsection{Review of the model\label{sec:review}} 

In this section we present the phenomenological model to describe the 1D flux power spectrum. Ultimately, we want to model the transmission fraction $F = \exp(-\tau)$, that is a function of the optical depth for Lyman-$\alpha$ photons, $\tau$. The fluctuations in the transmission fraction
are given  by
\be
\delta_F = \frac{F}{\bar F} - 1 \, ,
\label{deltaF}
\ee
where $\bar F$ denotes the average transmission fraction.
The transmission fraction, on the other hand, depends on the density contrast $\delta$ and the dimensionless gradient of the peculiar velocity $v_p$ along the line of sight
\be
\eta = - \frac{1}{aH} \frac{\partial v_p}{\partial x_p} \, ,
\ee
where $x_p$ is the comoving coordinate. At linear order, this relationship can be written as
\be
\delta_F = b_{F\delta} \, \delta + b_{F\eta} \, \eta \, ,
\ee
in terms of the  density contrast $\delta$ and the velocity gradient $\eta$.
In Zel'dovich approximation \cite{Zeldovich:1969sb}, the gradient of the velocity is 
proportional to the density contrast $\eta \propto \mu^2 \delta$, where $\mu$ is the angle between the line-of-sight and the momentum mode under consideration, $\mu = k_\parallel/k$ and $k_\parallel$ is the projection of the wavevector along the line-of-sight. Hence, the linear approximation to the three-dimensional flux power spectrum fulfills the relation
\be
P_F^{\rm linear}(k,k_\parallel,z) = b_{F\delta}^2 (1 + \beta \mu^2)^2 P_{\rm lin}(k,z) \, ,
\ee
where we introduced the parameter $\beta$ that parametrizes the proportionality between $\delta$ and $\eta$ and 
$P_{\rm lin}$ is the linear density power spectrum. Even though $\beta$ can be calculated in Zel'dovich approximation~\cite{Hui:1996fh}, it is known from simulations that this is not very accurate for the reionized intergalactic medium~\cite{McDonald:2001fe}. 
Hence, we keep $\beta$ as a free parameter in our setup.
In the final model, we eventually do not use the linear power spectrum but some theoretical models that reflect the clustering of the dark matter on short scales. At this point it also is possible to differentiate between the density and the velocity power spectra (see discussion below). 

In order to improve the model, we include two more physical effects.
The first one is the Jeans instability. 
Below the Jeans scale $k_J= a H /c_s$, baryonic density fluctuations cannot collapse. The Jeans scale is given in terms of the sound velocity 
\be
c_s^2 = \frac{T \gamma}{\mu_p m_p} \, ,
\ee
where $\mu_p m_p$  is the mean particle mass in the intergalactic medium  (we use $\mu_p \simeq 0.6$), $T$ its temperature, and $\gamma$ its adiabatic index. The bias function $b_{F\delta}^2$ is then modified by an additional suppression factor $\exp(-(k/k_F)^2)$ to account for the lack of clustering. The filtering scale $k_F$ can be understood as the redshift space average of the Jeans scale as proposed by Gnedin and Hui~\cite{Gnedin:1997td}
\be
\label{eq:defkJ}
\frac{1}{k_F(t)^2} = \frac{1}{D(t)} 
\int_0^t dt^\prime \, \frac{a^2(t^\prime)}{k^2_J(t^\prime)} 
\left[\frac{d}{dt^\prime} \left( a(t^\prime)^2  \, \frac{d}{dt^\prime} D(t^\prime) \right) \right]
\int_{t^\prime}^t \frac{dt^{\prime\prime}}{a^2(t^{\prime\prime})} \, .
\ee
There are further effects that can suppress the observed power along the line-of-sight $k_\parallel$. For example 
redshift space distortions due to peculiar velocities~\cite{Scoccimarro:2004tg}, the finite resolution of the experimental observation or thermal broadening~\cite{Hui:1997dp}. The most important effect is hereby the thermal broadening with the scale $k_s \simeq \sqrt{m_p/T}$ and we take these effects into account with another exponential suppression factor $\propto\exp(-k_\parallel^2/k_s^2)$~\cite{Hui:1997dp}.
In contrast to the analysis in \cite{Garny:2018byk}, we actually use $k_F$ and $k_s$ as model parameters rather than $T$ and $c_s$, since we want to be agnostic about the physics of the intergalactic medium. We explicitly checked that the redshift-dependence implied by (\ref{eq:defkJ}) has a negligible impact on the outcome compared to choosing a fixed value for $k_F$. The main reason is the overall mild effect of $k_F$ within our setup, as will be explained in detail below.

Finally,  a visible modulation in the observed flux power spectrum is induced by Si\,{\sc\small III} absorption. We model this effect with an additional factor 
\be
\kappa_{\rm SiIII} = 1  + 2 \left(\frac{\rm f_{SiIII}}{1 - \bar F}\right) \cos(\Delta V  \, k_\parallel) + 
\lp \frac{\rm f_{SiIII}}{1 - \bar F} \rp^2 \, ,
\label{SiIII}
\ee
following the literature~\cite{McDonald:2004eu, Palanque-Delabrouille:2013gaa}.
Here, $\bar F$ denotes the mean transmission fraction and we introduced the two parameters $\Delta V$ and ${\rm f_{SiIII}}$.

Integrating the three-dimensional flux power spectrum across the line-of-sight then yields the one-dimensional flux power spectrum~\cite{Kaiser:1990xe}, 
\be
P_{\rm 1D}(k_\parallel,z) = \frac{1}{2\pi}\int_{k_\parallel} \, P_F(k,k_\parallel,z) \, k \, dk \, ,
\ee
and we obtain
\be\label{eq:model}
P_{\rm 1D}(k_\parallel,z) = 
A \, \kappa_{\rm SiIII}(k_\parallel,z) \, (\log \bar F(z))^2 \, 
\exp( - (k_\parallel/k_s(z))^2) \, ( I_0 + 2 \beta(z) I_2 + \beta(z)^2 I_4 ) \, , 
\ee
where the parameter $A$ denotes the overall amplitude, and we defined the three integrals
\bea\label{eq:I024}
I_0(k_\parallel,z) &=& \int_{k_\parallel} \, dk \,k 
\, \exp(-(k/k_F)^2) \,  P_{\delta\delta}(k,z) + \bar I_0(z) \, , \nn \\
I_2(k_\parallel,z) &=& \int_{k_\parallel} \,  \frac{dk \,k_\parallel^2}{k} 
\, \exp(-(k/k_F)^2) \,  P_{\delta\theta}(k,z) \, , \nn \\
I_4(k_\parallel,z) &=& \int_{k_\parallel} \, \frac{dk \,k_\parallel^4}{k^3} 
\, \exp(-(k/k_F)^2) \,  P_{\theta\theta}(k,z) \, . 
\eea

As mentioned before, this result depends on the different power spectra and the cross-correlation:
$P_{\delta\delta}(k,z)$ is the density power spectrum, $P_{\theta\theta}(k,z)$ the power spectrum of the velocity divergence  and $P_{\delta\theta}(k)$ is the cross correlation with $\theta=-\nabla\vec v/(aHf)$~\footnote{The normalization in our definition of $\theta$ ensures that all power spectra are the same at the linear level. The growth rate $f$ does not appear explicitly below, because we absorb it into the definition of the velocity bias parameter $\beta$.}. 

Out of the three integrals $I_0, I_2, I_4$ the first one is somewhat special. The latter two are dominated by large-scale (IR) contributions $k \simeq k_\parallel$, while $I_0$ can have sizeable contributions from small (UV) scales. The main reason for this difference is the different weight of factors of $k$ in the integrand in \eqref{eq:I024}. In addition, the velocity and cross power spectra are smaller compared to the density on small scales \cite{Hahn:2014lca}. For that reason, we introduced a counter term $\bar I_0(z)$ that accounts for the unknown UV contributions.

The redshift dependence of the parameters $\bar F(z)$, $\beta(z)$ and $\bar I_0(z)$ is not specified so far. In the simplest model, we use a polynomial dependence on the scale factor, i.e.
\bea\label{eq:alphabeta}
\beta &=& \alpha_{\rm bias} \, [a(z_{\rm pivot})/a(z)]^{\beta_{\rm bias}}, \nn \\
\log \bar F  &=& \alpha_F \, [a(z_{\rm pivot})/a(z)]^{\beta_F} , \nn \\ 
\bar I_0 &=& \alpha_{\rm c.t.} \, [a(z)/a(z_{\rm pivot})]^{\beta_{\rm c.t.}}\,.
\eea
We use $z_{\rm pivot}=3$. We note that the effect of He\,{\small\sc II} reionization on the IGM temperature and adiabatic index is often taken into account by a broken power law ansatz. Since the IGM properties are described by the counterterm and bias parameters within the effective model, this could motivate an extension of \eqref{eq:alphabeta} to a broken power law. However, as we focus on a rather narrow redshift range in this work ($3\leq z\leq 4.2$, see below), the approximation of the redshift-dependence with a single power law is sufficient. In addition, analyses of the IGM state \cite{Becker_2010,Gaikwad:2020eip} suggest that the impact of He\,{\small\sc II} reionization either occurs rapidly for $z\lesssim 3$, i.e. outside the range of interest to us, or more smoothly at higher redshifts, in which case a single power law provides a reasonable description for  $3\leq z\leq 4.2$ as well.

As we will see below, the effective model can provide an accurate description of both simulation and observation data. Nevertheless, in principle, the model could be further extended, taking for example additional higher-order bias parameters into account~\cite{Desjacques:2018pfv}. We also note that in the analysis of BOSS Lyman-$\alpha$ data performed in \cite{Palanque-Delabrouille:2015pga,Palanque-Delabrouille:2019iyz}, the impact of several astrophysical and systematic uncertainties has been accounted for by various multiplicative and additive terms with coefficients treated as nuisance parameters.
While some of these uncertainties are not present in our analysis (e.g. related to combining hydrodynamical simulations of different resolution), the extra free parameters considered in \cite{Palanque-Delabrouille:2015pga} that are related to residual contamination from damped Lyman-$\alpha$ (DLA) systems, from astrophysical feedback processes, as well as UV background fluctuations could be introduced in the same way within the effective model as well. However, their effect is largely degenerate with the free parameters that were introduced already. While there is no simple one-to-one correspondence in general, the parameterization to account for UV background fluctuations adopted in \cite{Palanque-Delabrouille:2015pga,Palanque-Delabrouille:2019iyz} can for example be absorbed in the counterterm parameters within our model. The DLA and feedback corrections corresponds to a combination of counterterm, bias and $k_s$ parameters. Therefore, we do not explicitly include the extra nuisance parameters considered in \cite{Palanque-Delabrouille:2015pga,Palanque-Delabrouille:2019iyz} in our analysis.

In total, our model then incorporates eleven parameters,
\be
\label{eq:freeparams}
\{ A,\, \alpha_{F},\,  \beta_{F},\, \Delta V,\, f_{\rm SiIII},\, k_s,\, k_F,\,  \alpha_{\rm bias},\, \beta_{\rm bias},\,\alpha_{\rm c.t.},\, \beta_{\rm c.t.}  \}\,.
\ee
In principle, all these parameters can be determined from simulations of the intergalactic medium and all parameters carry uncertainties 
due to our limited knowledge about the dynamics of the IGM.
However, not all parameters are always relevant. For relatively small wavenumbers within the range measured by BOSS ($k\leq 0.02($km/s$)^{-1} \sim 2$h/Mpc), the Jeans scale and the thermal broadening are not essential. This means that the choice of $k_F$ and $k_s$ has only a mild impact on the model. Instead, the relevant parameters absorbing the dependence on the IGM properties are the bias as well as the counterterm.
Moreover, when we fit to hydrodynamic simulations, the power spectrum does not contain modulations from Si\,{\sc\small III}. In this case, we do not need the parameters $\Delta V$ and $f_{\rm SiIII}$. 

For our baseline analysis we fix all parameters that have only a minor impact. In particular, we find that neglecting the redshift dependence of $k_F$ within
the range we consider has a minor impact, setting $k_F=18\,$h/Mpc, and using $k_s=0.11 ({\rm km}/{\rm s})^{-1} \simeq 13\,$h/Mpc \cite{Garny:2018byk}. 
In addition, we fix the parameters related to Si\,{\sc\small III} cross correlation, that are well constrained by
BOSS data \cite{Palanque-Delabrouille:2013gaa,Chabanier:2018rga}. Specifically, we use fiducial values for the transmission fraction entering in Eq.\,\eqref{SiIII}, effectively replacing $f_{\rm SiIII}/(1-\bar F)\to f_{\rm SiIII}/(1-\bar F_{\rm fid})$
with $f_{\rm SiIII}=6\cdot 10^{-3}$, $\Delta V=2\pi/0.0028$, $\bar F_{\rm fid}=\exp(-0.0025(1+z)^{3.7})$. 
Note that, within this approximation, the parameter $\beta_F$ entering
the factor $\ln\bar F(z)$ in Eq.\,\eqref{eq:model} is kept as a free parameter, while $\alpha_{F}$ can be absorbed into $A$. In summary, the free parameters in our baseline model are
\be\label{eq:baseline}
  \{ A,\, \beta_{F},\, \alpha_{\rm bias},\, \beta_{\rm bias},\,\alpha_{\rm c.t.},\, \beta_{\rm c.t.} \}.
\ee
We use no priors on these parameters in our fiducial analysis, that can therefore be considered as rather conservative. 
After validating the baseline model by comparing to hydrodynamical simulations, we also consider various modifications to assess the robustness and predictivity.

\subsection{Input power spectra for massive neutrinos} 

After the ultra-relativistic neutrinos decouple from the thermal bath in the early Universe, their large thermal velocities lead to
a strong suppression of their density contrast compared to baryons and cold dark matter~\cite{Lesgourgues:2006nd}. The wavenumber above which this suppression sets in
is given by the comoving free-streaming scale~\cite{Shoji:2010}
\be
 k_{\rm fs} \simeq \frac{0.0908}{(1+z)^{1/2}}\frac{m_\nu}{0.1\,{\rm eV}}\sqrt{\Omega_m^0}h/{\rm Mpc}\,,
\ee
where $m_\nu$ is the neutrino mass and $\Omega_m^0$ is the matter density parameter today.

For $z\ll z_{\rm nr}=189\,m_\nu/0.1\,$eV the neutrino background does contribute as a non-relativistic matter species to the expansion rate, while free-streaming
suppresses its contribution to the perturbations of the metric, i.e.~the gravitational potential. This leads to a slowdown of the growth rate
of baryon and cold dark matter perturbations for $k\gtrsim k_{\rm fs}$ as compared to the case without massive neutrinos, and correspondingly
to a scale- and redshift-dependent suppression of the linear matter power spectrum. At $z=0$ and for $k\gg k_{\rm fs}$ the relative suppression
asymptotes to approximately $-8f_\nu$ for small neutrino masses, where
\be
  f_\nu = \frac{\Omega_\nu}{\Omega_m} \simeq \frac{1}{\Omega_m^0h^2}\frac{\sum m_\nu}{93.14\,{\rm eV}}\,,
\ee
is the neutrino fraction.

Within the weakly non-linear regime, the suppression of the power spectrum is even more pronounced. Furthermore, the scale-dependence of the
suppression relative to a cosmological model with massless neutrinos extends to smaller scales when taking non-linearities into
account. At even smaller scales, this effect turns around \cite{Bird:2011rb,Hannestad:2020rzl}, 
known as ``spoon''-effect. Within the halo model, the dip of the spoon can be associated with the scale at which the non-perturbative 1-halo term overtakes the 2-halo contribution~\cite{Massara:2014kba,Hannestad:2020rzl}. While the dip of the spoon is at around $1h/$Mpc at $z=0$, it occurs beyond $5h/$Mpc for $z\gtrsim 3$. Therefore, the turnaround occurs on scales that are smaller than those covered by BOSS Lyman-$\alpha$ observations. Within the effective model, any spoon features are therefore completely absorbed into the counterterm parameters. In our approach, it is sufficient to control the three-dimensional input power spectrum within the BOSS range, $k\lesssim 2h/$Mpc. Since, at $z\gtrsim 3$, non-linearities in the matter density are much weaker than at $z=0$, perturbation theory methods can be used for the matter power spectrum down to smaller scales \cite{Garny:2018byk}. Indeed, the non-linear scale moves from $k\sim 0.3h/$Mpc at $z=0$ to $k\sim 2h/$Mpc at $z\gtrsim 3$.
For $z\gtrsim 3$ and $k\lesssim 2 h$/Mpc, relevant for the BOSS Lyman-$\alpha$ observations, non-linear corrections to the three-dimensional
matter-, velocity- and cross power spectra, that are required as an input for the effective Lyman-$\alpha$ model described above, can therefore be
estimated based on perturbative techniques \cite{Garny:2018byk}.
In our analysis, the power spectra $P_{\delta\delta}$, $P_{\delta\theta}$ and $P_{\theta\theta}$ entering \eqref{eq:I024} are computed at 1-loop.
For massive neutrinos, we compute the 1-loop contribution using the linear power spectrum of the combined cold dark matter and baryon perturbations $P_{\rm lin}^{cb,cb}(k,z)$ as input, as proposed in
\cite{Castorina:2015bma}, and add it to the total linear matter power spectrum $P_{\rm lin}$,
\be\label{eq:P1loop}
  P_{\delta\theta}(k,z) = P_{\rm lin}(k,z) + (1-f_\nu)^2P_{\rm 1-loop}^{\delta\theta}(k,z;P_{\rm lin}^{cb,cb})\,,
\ee
where 
\bea
  P_{\rm 1-loop}^{\delta\theta}(k,z;P_0) &\equiv& \int d^3q \Big[ 2F_2(k,q-k)G_2(k,q-k)P_0(|q-k|,z) \nn\\
  && {} +3(F_3(k,q,-q)+G_3(k,q,-q))P_0(k,z)\Big]P_0(q,z)\,.
\eea
Here $F_n$ and $G_n$ denote the usual perturbation theory kernels for the density and velocity divergence, respectively \cite{Bernardeau:2001qr}. For the $\delta\delta$ power spectrum
one replaces $G_n\to F_n$, and vice versa for $\theta\theta$. The linear power spectra are computed using the Boltzmann solver CLASS \cite{Blas:2011rf}.

The rationale for computing the loop correction with the cb part of the power spectrum only is that free-streaming suppressed the contribution from neutrino perturbations
on scales where non-linearities become important. This prescription has been tested against N-body simulation results for the total matter power spectrum in \cite{Castorina:2015bma}.
The impact of using the conventional perturbation theory kernels for massive neutrino cosmologies has recently been scrutinized by comparing to numerically evolved kernels, fully taking the
redshift- and scale-dependent growth in presence of massive neutrinos into account at the non-linear level \cite{Garny:2020ilv}. While, at $z=0$, percent-level deviations have been found, the approximation scheme based on computing non-linear corrections for the cold dark matter and baryon perturbations with standard kernels as described above was found to work well far above the neutrino free-streaming scale, and within
the matter dominated era \cite{Garny:2020ilv}, which is the case for Lyman-$\alpha$ observations. Therefore, we adopt this ``cb'' approach here. Nevertheless, it may be interesting to investigate the impact of a relative velocity difference of the cold dark matter and baryon components at $z\gtrsim 3$ in the future \cite{Bird:2020kwe,Fernandez:2020jgf}. In addition, in \cite{Villaescusa-Navarro:2017mfx} it was argued that the distribution of halos follows the cb component even on large scales. We checked that using the linear cb spectrum in~\eqref{eq:P1loop} (as well as in the analogous expressions for $P_{\delta\delta}$ and $P_{\theta\theta}$) instead of the full matter
power spectrum has only a minor impact on our results (see discussion at the end of section~\ref{sec:robust}).

\begin{figure}[t!]
        \centering
        \includegraphics[width=0.65\textwidth]{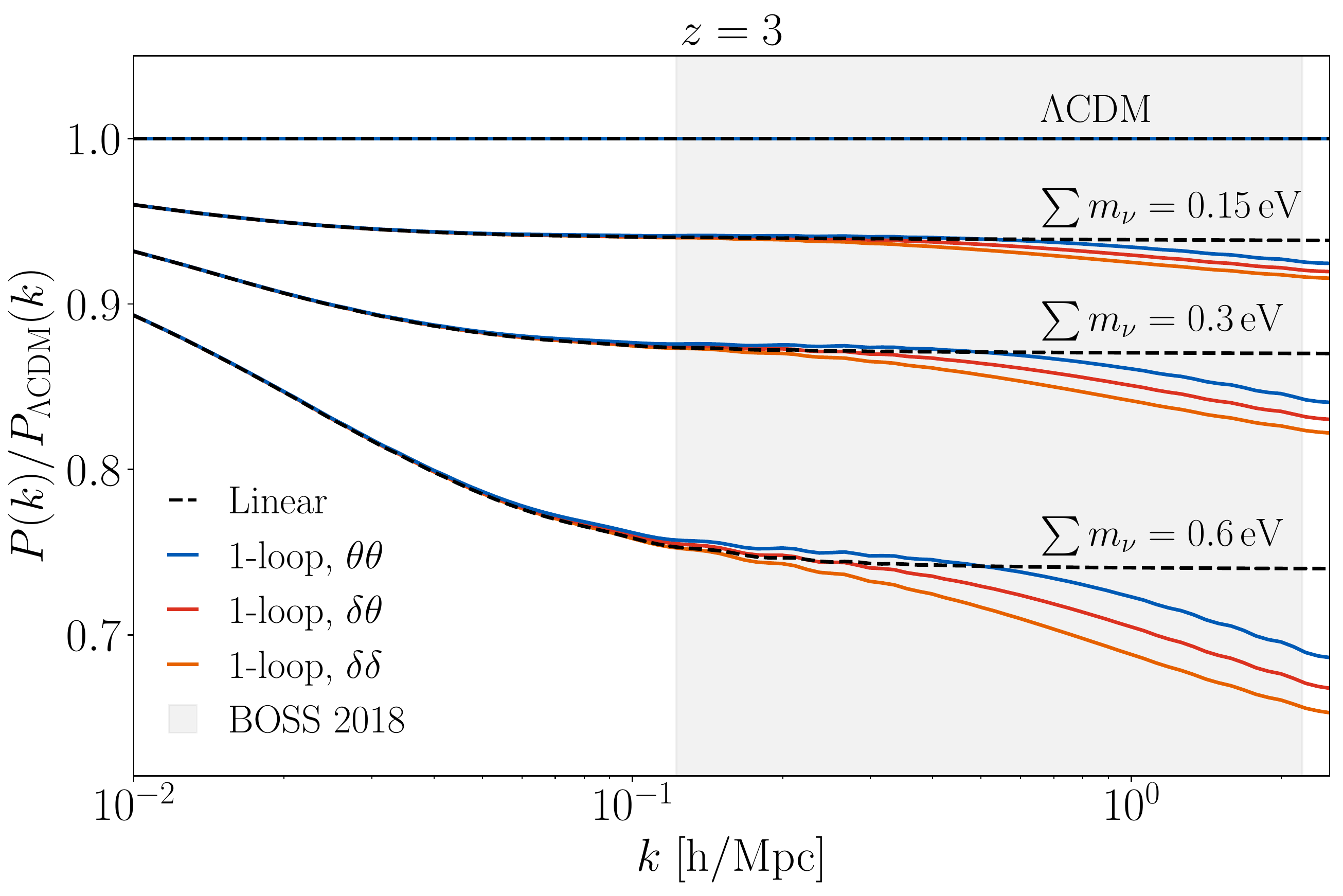}
        \caption{Matter power spectrum for massive neutrinos for $\sum m_\nu=0,0.15,0.3,0.6$\,eV and $z=3$ normalized to the $\Lambda$CDM spectrum. Dashed lines show the linear power spectrum and solid lines the 1-loop results for $P_{\theta\theta}$ (blue), $P_{\delta\theta}$ (red) and $P_{\delta\delta}$ (orange),  respectively. The gray shaded region indicates the scales of BOSS observations.}
        \label{fig:input_powerspec}
\end{figure}

In figure~\ref{fig:input_powerspec}, we show the ratio of the matter power spectra for massive neutrinos at $z=3$, relative to the massless case. While the ratio of
linear power spectra asymptotes to a plateau (that is slightly smaller than $-8f_\nu$ at $z=3$), 
the 1-loop power spectrum shows a further scale-dependent suppression on weakly non-linear scales. In addition,
while the density and velocity power spectra agree at the linear level for the normalization of $\theta$ adopted here, they differ at the 1-loop level, with a
stronger suppression for the \emph{ratio} $P_{\delta\delta}/P^{\Lambda{\rm CDM}}_{\delta\delta}$ as compared to $P_{\theta\theta}/P^{\Lambda{\rm CDM}}_{\theta\theta}$ (note that $P_{\theta\theta}/P_{\delta\delta}\leq 1$ for a fixed neutrino mass). 

Within the range of the BOSS Lyman-$\alpha$ data ($k\sim 0.1-2\,h/$Mpc), the suppression of the \emph{linear} matter power spectrum is almost scale-independent
for $\sum m_\nu\lesssim 0.5\,$eV. The additional scale-dependence of the power spectrum due to non-linear corrections is therefore an important feature for
probing the neutrino mass with Lyman-$\alpha$ observations. In addition, the dependence of the power spectrum on redshift is sensitive to the neutrino mass.
For the scales and redshifts $4\geq z\geq 3$ considered here, it can be approximately described by
\be
  P_{\delta\delta}(k,z) = D_{\rm eff}(z)^2P_{\rm lin}(k,z_{\rm pivot}) + D_{\rm eff}(z)^4(1-f_\nu)^2P_{\rm 1-loop}^{\delta\delta}(k,z_{\rm pivot};P_{\rm lin}^{cb,cb})\,,
\ee
(and analogously for $\delta\theta$ and $\theta\theta$) where 
\be
  D_{\rm eff}(z)=\left(\frac{1+z_{\rm pivot}}{1+z}\right)^{1-3f_\nu/5}\,,
\ee
is the growth factor appropriate for matter domination and $k\gg k_{\rm fs}$.

Due to the almost scale-independent suppression of the linear power spectrum on scales relevant for Lyman-$\alpha$ observations, it is possible to approximately ``cancel'' the suppression
by increasing the normalization of the primordial power spectrum, described by the parameter $A_s$ within $\Lambda$CDM. It has been stressed in \cite{Pedersen:2019ieb}
that this leads to a degeneracy between the sum of neutrino masses $\sum m_\nu$ and $A_s$. In this work we fix the value of $A_s$ when comparing models with different neutrino masses,
motivated by the strong constraints on this parameter from Planck~\cite{Aghanim:2018eyx}, thereby breaking the degeneracy. Nevertheless, as we will see, a similar
degeneracy occurs when using the most conservative realization of the effective model for the 1D Lyman-$\alpha$ power spectrum with a completely free amplitude $A$ in \eqref{eq:model}.
After validating the baseline model for the Lyman-$\alpha$ power spectrum, we will therefore discuss to which extent the parameter $A$ can be restricted by comparing to
hydrodynamical simulations.

\section{Validation with simulation data} \label{sec:val}

\subsection{Fit of the effective model to simulation data} 

In order to validate our effective model for the one-dimensional Lyman-$\alpha$ flux power spectrum,
we compare to hydrodynamical simulation data \cite{bolton17}. The simulations are based on a $\Lambda$CDM
cosmology with $h=0.678$, $\Omega_b=0.0482$, $\Omega_{\rm CDM}=0.260$, $n_s=0.961$, $A_s=2.12\cdot 10^{-9}$, $\tau=0.0952$,
and varying values for the sum of neutrino masses $\sum m_\nu=0,0.15,0.3,0.6,0.9$\,eV.
Note that we keep the baryon and cold dark matter density parameters fixed for all cases, for the purpose
of comparing simulations with theoretical predictions of the effective model.

The neutrino simulations are based on the particle implementation described in \cite{viel10}. Basically, neutrinos are simulated with an extra set of particles implemented in the initial conditions of the hydrodynamic simulations with the correct thermal velocities. Even if other approaches can also be used to simulate neutrino non-linear clustering, the particle based approach guarantees the most accurate non-linear behaviour, once neutrino shot-noise is under control.
The simulations are performed in a cubic box with comoving side length $40h/$Mpc, and using $512^3$ particles for neutrinos, cold dark matter, and baryons, respectively.
Furthermore, the IGM parameters correspond to the reference case adopted in \cite{bolton17}.
We consider wavenumbers within the BOSS range $0.001-0.02($km/s$)^{-1}\sim 0.1-2\,h/$Mpc, and focus on redshifts $z=3.0, 3.2, 3.4, 3.6, 3.8, 4.0, 4.2$.
While the BOSS data \cite{Chabanier:2018rga} encompass redshifts within the range $2.2-4.6$, we conservatively omit redshifts below $3.0$ since
they are more strongly affected by non-linearities on the scales observed by BOSS, beyond the validity of the analytical model,
as well as redshifts above $4.2$ due to the increased sensitivity to reionization physics. Furthermore, in order to assess the potential of constraining the neutrino mass
with BOSS data we assign relative errors to the
simulation data that are equal to those quoted by BOSS \cite{Chabanier:2018rga} (see section~\ref{sec:boss} for details) and then perform a $\chi^2$ fit.

\medskip

\noindent We consider two models for the input power spectra $P_{\delta\delta}$, $P_{\delta\theta}$ and $P_{\theta\theta}$:
\begin{itemize}
\item[$(i)$] using the linear power spectrum, and 
\item[$(ii)$] 1-loop power spectra.
\end{itemize}
Note that, due to the counterterm included in \eqref{eq:I024}, even $(i)$ captures the impact of strongly non-linear effects on small scales on the 1D power spectrum to a certain degree.
The main difference between $(i)$ and $(ii)$ is the redshift- and scale dependence \emph{within} the BOSS range, corresponding to weakly non-linear scales.

\begin{figure}[t!]
        \centering
        \includegraphics[width=\textwidth]{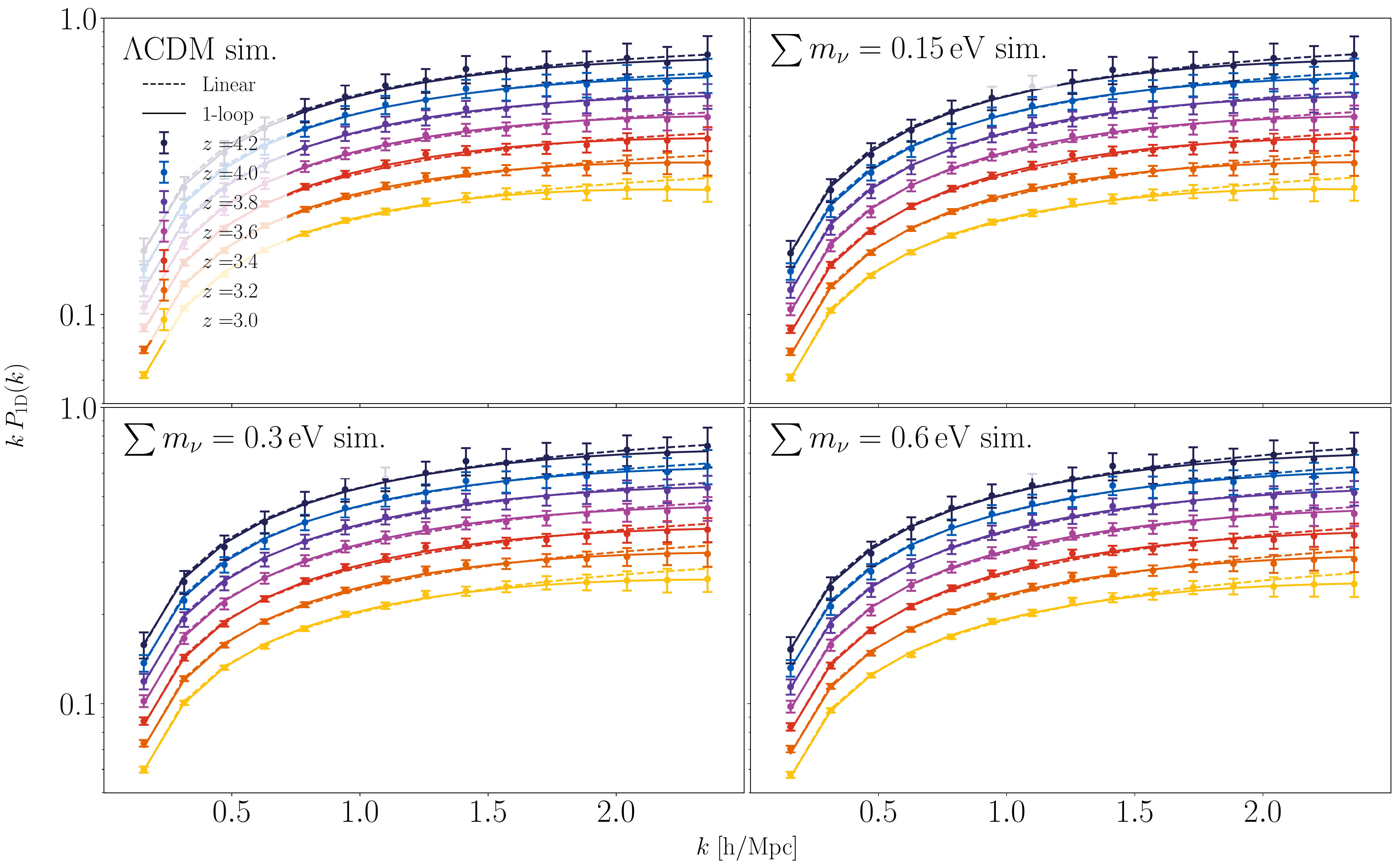}
        \caption{1D Lyman-$\alpha$ flux power spectrum from hydrodynamical simulations (data points) and best fit analytical model when using 1-loop power spectra (solid lines) or the linear power spectrum (dashed lines) as input, for $\sum m_\nu=0,0.15,0.3,0.6$\,eV.}
        \label{fig:pk_sims}
\end{figure}

In figure\,\ref{fig:pk_sims} we show the best fit analytical model for the one-dimensional flux power spectrum, compared to the
simulation data, for $\sum m_\nu=0,0.15,0.3,0.6$\,eV. We observe that the analytical model with 1-loop input power spectrum
can describe the simulation data well for all redshifts and neutrino masses. Since the simulations are not affected by observational
errors, the total value of $\chi^2$ is rather low when using the error bars taken from BOSS \cite{Chabanier:2018rga}
($\chi^2=10.84,12.17,14.90,19.32$ for $\sum m_\nu=0,0.15,0.3,0.6$\,eV, respectively, with $99$ degrees of freedom). A similar behaviour
has been noted in \cite{Arinyo-i-Prats:2015vqa}.  Below we will see that the total $\chi^2$ value is of the order of the number of degrees of freedom
when fitting to the BOSS data. For the simulations, our total value of $\chi^2$ should therefore not be understood in absolute terms as a real ``goodness of fit", but rather its variation reflects whether the model fit improves or not. When using the linear power spectra as input for the analytical model instead of the
1-loop approximation, the fit to the simulation is slightly worse, showing a systematic overestimation of the power towards high $k$, and correspondingly
higher values $\chi^2=32.55,31.97,33.13,37.65$.

We checked that the agreement between the analytical model and the simulation data
does not depend on the cutoff that is used for the numerical evaluation of the integrals in \eqref{eq:I024} (we use $20\,h/$Mpc as default value,
and checked that using $10\,h/$Mpc instead does not influence our results). While, as expected, the value of
$I_0$ does depend on the cutoff, this dependence is absorbed in a shift of the model parameters for the flux power spectrum, in particular the counterterm parameter
$\alpha_{\rm c.t.}$. Furthermore, our results are stable against variations of the parameters $k_s$ and $k_F$; we will come back to this point in section~\ref{sec:boss}.
We conclude that the dominant impact of the complex physics of the intergalactic medium can be accounted for by the free parameters of the analytical model.

\subsection{Discrimination of $\Lambda$CDM vs massive neutrinos} 

\begin{figure}[t!]
        \centering
        \includegraphics[width=\textwidth]{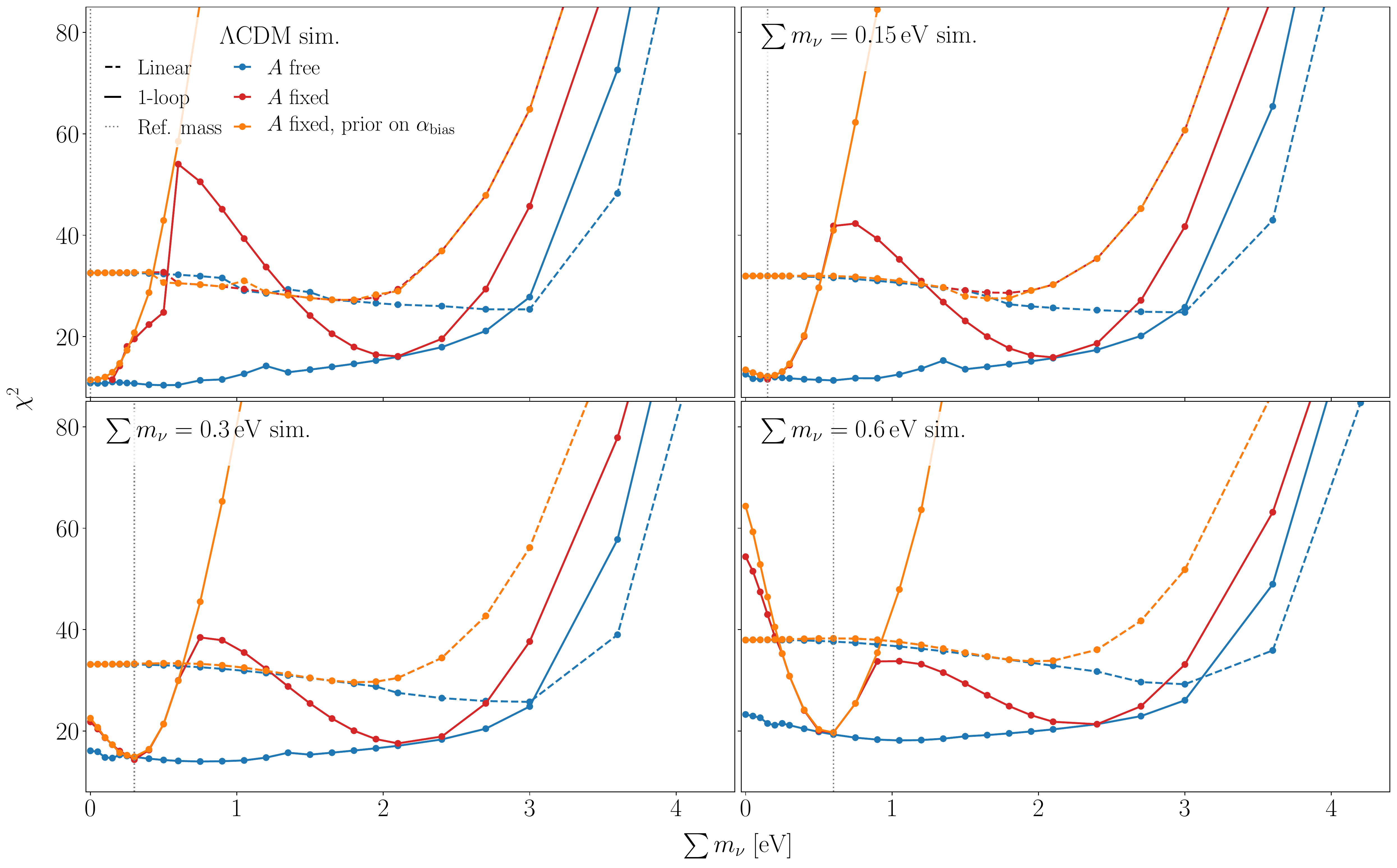}\\
        \caption{$\chi^2$ obtained from a fit of the analytical model to hydrodynamical simulations. The panels correspond to four values of
the ``true'' neutrino mass ($\sum m_\nu=0,0.15,0.3,0.6$\,eV) used in the simulation. Blue lines show the $\chi^2$ obtained when leaving all parameters of the baseline analytical model free, while the amplitude $A$ is fixed according to \eqref{eq:A} for the red lines. Imposing in addition a $50\%$ prior on $\alpha_{\rm bias}$ yields the orange curves. In all cases solid lines correspond to 1-loop, and dashed to linear input power spectra.}
        \label{fig:chi2_sims}
\end{figure}

In order to investigate in how far the analytical model for the Lyman-$\alpha$ flux power spectrum can be used to
set constraints on the sum of neutrino masses, we fit the hydrodynamical simulation data 
for a set of cosmological models with varying ``input'' neutrino mass, that does not necessarily match the ``true'' value of $\sum m_\nu$ of the simulation. 
The resulting $\chi^2$ values are shown in figure \ref{fig:chi2_sims} 
as a function of the ``input'' mass, and for the four simulations corresponding to a ``true'' neutrino mass $\sum m_\nu=0,0.15,0.3,0.6$\,eV, respectively. 
As mentioned before, the absolute value of $\chi^2$ should be regarded with care when fitting to
simulation data. Nevertheless, we take the relative differences as an indicator of the sensitivity to the neutrino mass (we find comparable
differences in $\chi^2$ for the BOSS data, see below). When using the baseline Lyman-$\alpha$ model (blue solid line), the sensitivity to the sum of neutrino
masses is rather weak. In other words, the analytical model can describe the simulated data very well for a large range of ``input'' neutrino masses.
The main reason is that we leave the overall amplitude $A$ completely free in the baseline model. Together with shifts in the other free
parameters, a change in $A$ can compensate for the suppression of the power spectrum depending on the neutrino mass. This degeneracy is not
perfect however, due to the different redshift- and scale-dependence of the 1-loop power spectrum depending on the neutrino mass.
In particular, for values of $\sum m_\nu$ of order eV or larger, the free-streaming scale moves towards the range of wavenumbers observed by BOSS, such that
the suppression of the power spectrum becomes strongly scale-dependent within the BOSS window, thereby breaking the degeneracy with $A$.
This is also apparent when comparing to the analytical model based on the linear power spectra (blue dashed lines). While, as observed above,
the overall $\chi^2$ is much larger in that case, the sensitivity to the neutrino mass is even smaller. This is expected, because in this case
the scale-dependence of the input power spectrum is less pronounced within the relevant range of wavenumbers, unless for extremely
large neutrino masses, where the free-streaming scale approaches the BOSS range (see figure~\ref{fig:input_powerspec}).

\begin{figure}[t!]
        \centering
        \includegraphics[width=0.5\textwidth]{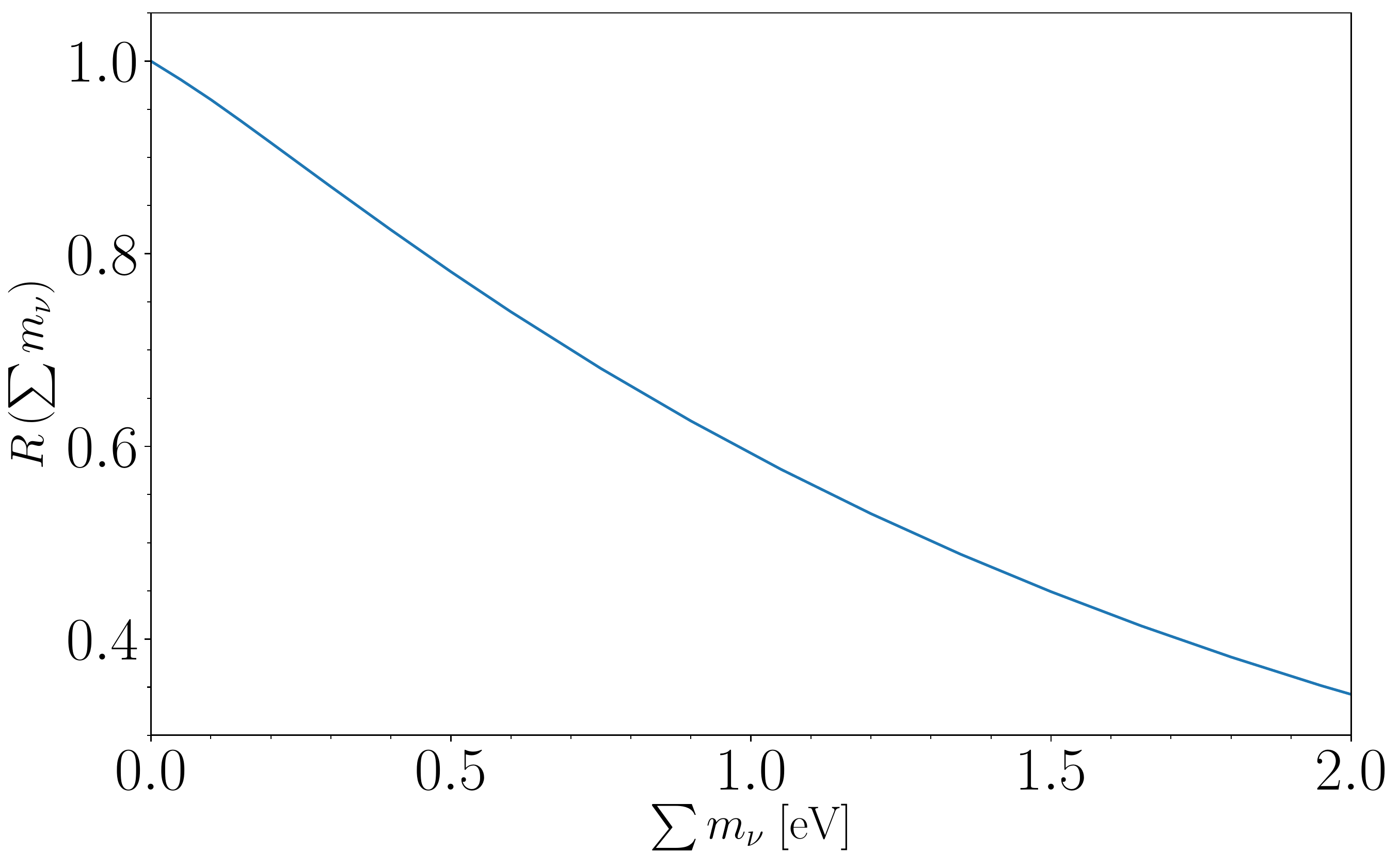}
        \caption{Ratio $R$ characterizing the suppression of the linear matter power spectrum relative to $\Lambda$CDM for $k\gg k_{\rm fs}$ and $z_{\rm ref}=3$.}
        \label{fig:Rvsmnu}
\end{figure}

We therefore also consider a restricted version of the analytical model for which the dependence of the amplitude $A$ on the neutrino mass
is assumed to be known. Specifically, we make the ansatz
\be\label{eq:A}
  A = A_0\times\frac{1}{R^c}
\ee
where $R$ is the plateau value of the linear matter power spectrum relative to the corresponding $\Lambda$CDM model with massless neutrinos,
\be
  R({\textstyle\sum} m_\nu) \equiv \frac{P_{lin}(k,z_{\rm ref};\sum m_\nu)}{P_{lin}(k,z_{\rm ref};0)}\Big|_{k\gg k_{\rm fs}}\,,
\ee
where we use $z_{\rm ref}=3$ (see figure \ref{fig:Rvsmnu}). The amplitude $A_0$ is set to the best-fit value determined from fitting the $\Lambda$CDM model with massless neutrinos to the
corresponding simulated Lyman-$\alpha$ flux power spectrum, and the power law index $c$ is calibrated by comparing to the hydrodynamical simulations for massive neutrinos.
We find that for $c=0.8$, the $\chi^2$ value when using the ansatz from above to fix the amplitude is the same as for the baseline model with free amplitude for
all neutrino masses $\sum m_\nu=0.15,0.3,0.6$\,eV, when using the value of the ``input'' neutrino mass in the fit that matches the ``true'' neutrino mass of the
simulation. We therefore adopt this choice for $c$ in the following. Note that for $c=1$, the rescaling of the amplitude would exactly compensate for the
suppression of the linear power spectrum. Since the mean Lyman-$\alpha$ flux obtained in the hydrodynamical simulations is rescaled to the observed mean
flux, one may expect $c$ to be close to one, up to corrections due to the scale- and redshift-dependence of the input power spectrum within the BOSS range.
In summary, the restricted model with $A$ fixed by \eqref{eq:A} performs as well as the baseline model when fitting models with an ``input'' value for
the neutrino mass that matches the ``true'' one used in the simulation. This is non-trivial since \eqref{eq:A} contains only two free parameters, while
we consider simulations for four sets of neutrino masses, covering a wide range. We also checked that the agreement extends to higher neutrino masses
using a simulation with $\sum m_\nu=0.9$\,eV.

In the next step, we reconsider the possibility to constrain the neutrino mass. We therefore fit again cosmological models with varying ``input'' neutrino mass to each
of the simulations. The corresponding $\chi^2$ values are shown by the red lines in figure \ref{fig:chi2_sims}. We observe that, when using 1-loop power spectra (red solid lines),
the $\chi^2$ function features a pronounced minimum at the ``true'' neutrino mass for all simulations. The observation that the value of $\chi^2$ at this minimum lies on
top of the blue line corresponds to the finding discussed in the previous paragraph. In addition, the red lines feature a second minimum at a significantly higher neutrino
mass. This feature can be attributed to a parameter degeneracy between the bias and counterterm parameters. Nevertheless, the value of $\chi^2$ at the second minimum is
larger than for the minimum at the ``true'' value. We find that, in practice, this feature does not impact the constraint on the sum of neutrino masses for realistic values
of its ``true'' value (see below). Nevertheless, we point out that the degeneracy can be broken by imposing in addition a weak prior on the bias parameter. In particular,
if we require that the bias parameter $\alpha_{\rm bias}$ lies within $\pm 50\%$ of the best-fit value for the $\Lambda$CDM model (that is, using the bias obtained for
vanishing neutrino mass in both the ``input'' and ``true'' value as reference value). The corresponding $\chi^2$ values with prior on the bias
are shown by the orange solid lines in figure~\ref{fig:chi2_sims}. While the (spurious) second minimum is lifted, we observe that the $\chi^2$ values around the minimum
at the ``true'' neutrino mass are robust. Finally, we remark that when using the linear power spectrum instead of 1-loop, there is no sensitivity to the neutrino mass even
when fixing the amplitude according to \eqref{eq:A} (red dashed lines) and imposing a prior on the bias (orange dashed lines). As expected, the additional scale-dependence of
the power spectrum due to non-linear corrections is crucial for being able to constrain the neutrino mass.

\begin{figure}[t!]
        \centering
        \includegraphics[width=0.65\textwidth]{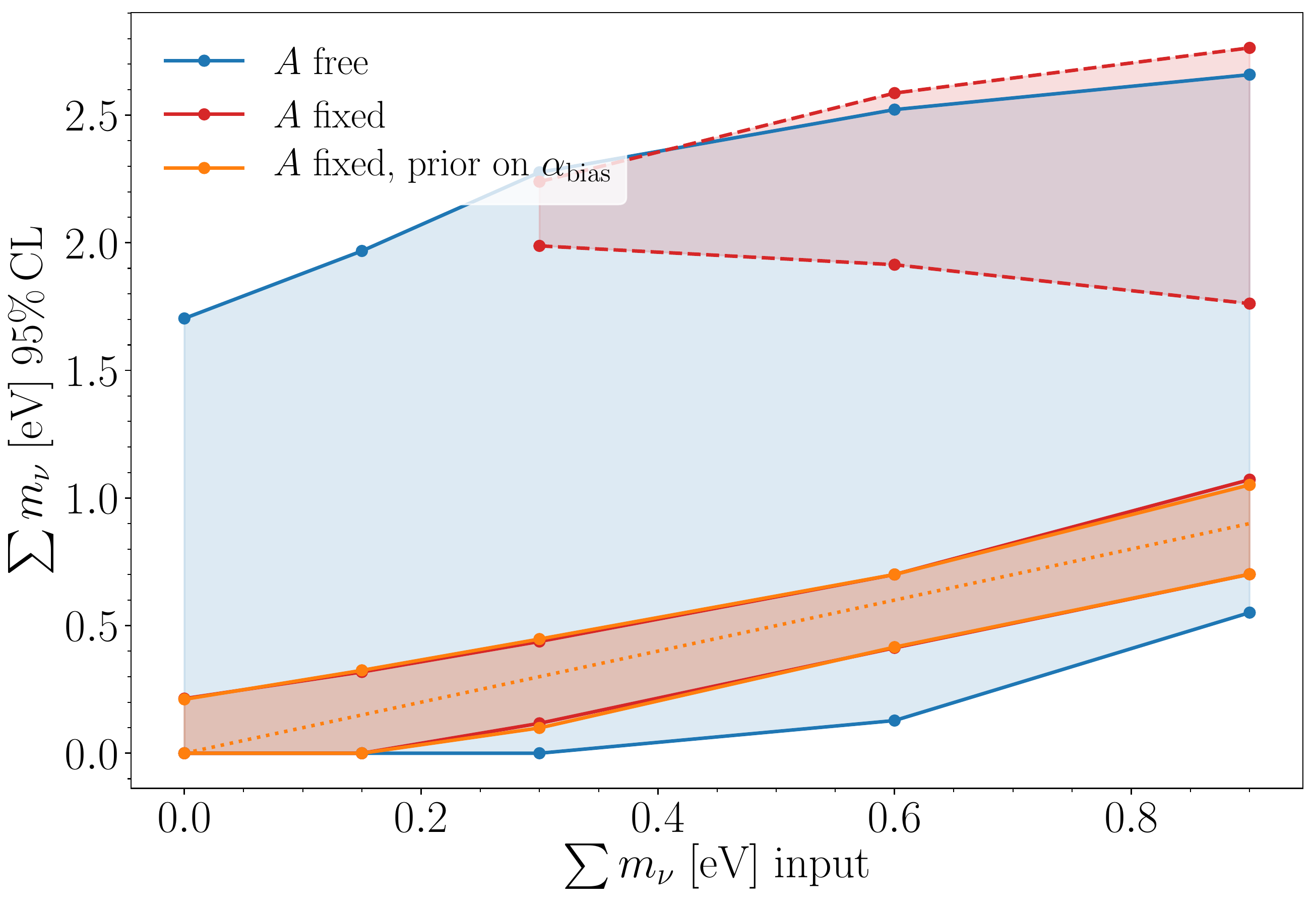}\\
        \caption{Inferred $95\%$ C.L. interval for $\sum m_\nu$ when fitting to simulation data with ``true'' neutrino mass given on the $x$-axis. The blue shaded region corresponds to the case where all parameters are left free. The red region is obtained for fixed amplitude, according to \eqref{eq:A}, and the orange region when assuming in addition a $50\%$ prior on $\alpha_{\rm bias}$. Note that the orange and red regions almost overlap. For the case of free bias, a second solution is obtained for large neutrino masses (red region with dashed lines).
}
        \label{fig:mnufit}
\end{figure}

In figure \ref{fig:mnufit} we show the $95\%$ C.L. interval for the neutrino mass determined from requiring that $\Delta\chi^2<3.84$ (corresponding to $p\leq 0.05$ for a $\chi^2_1$ distribution)  relative to the minimum value.
The horizontal axis corresponds to the ``true'' neutrino mass used in the simulation, while the vertical axis shows the resulting upper and lower bound.
Since we are assigning error bars to the simulation data that correspond to the ones of the measured flux power spectrum, the precision with which the
neutrino mass can be reconstructed should give a realistic estimate of the expected sensitivity. For the baseline model with all parameters left free,
the sensitivity is very weak, and no meaningful bound can be extracted (blue area in figure \ref{fig:mnufit}). When fixing the amplitude according to \eqref{eq:A}, one can infer an unbiased estimate
of the ``true'' neutrino mass with a precision of about $0.15\,$eV above or below the ``true'' value (red area). 
For ``true'' values larger than $0.15\,$eV, one also obtains a lower bound, i.e. can discriminate the neutrino mass from zero at $2\sigma$, while
for ``true'' values smaller than $0.1\,$eV, one expects an upper bound of $0.15-0.2\,$eV. For (unrealistically) large ``true'' values, neutrino masses in the vicinity of the second minimum are
also allowed at $2\sigma$ (red area with dashed boundary). This spurious region is eliminated when imposing a prior on the bias (orange shaded region). Note that, for realistic neutrino masses,
the bound extracted with or without prior on $\alpha_{\rm bias}$ is almost identical. We will therefore use the more conservative Lyman-$\alpha$ model with free bias in our analysis of the BOSS data.

We performed numerous additional tests in order to assess the robustness of the expected neutrino mass sensitivity. In particular, we checked that the cutoff used in the
integrals \eqref{eq:I024} has no effect on the result (we used values in the range $10-20$h/Mpc). While a change in the cutoff does lead to different best-fit parameters
(in particular for the counterterm, as expected), the shape of the $\chi^2$ curves is robust, in particular close to the minimum. We also verified that the $\chi^2$ curves
depend very weakly on the parameters $k_s$ and $k_F$. We will quantify their impact in more detail after presenting results obtained from the Lyman-$\alpha$ flux power spectrum
measured by BOSS in the next section.

\section{Application to BOSS data \label{sec:boss}} 

In this section, we apply the effective model for the 1D Lyman-$\alpha$ flux power spectrum to the data reported by BOSS~\cite{Chabanier:2018rga}\footnote{We use the 1D Lyman-$\alpha$ flux power spectrum as provided in the data attached to~\cite{Chabanier:2018rga}, in particular the third column of {\tt Pk1D\_data.dat}. For the error used in the fit we sum in quadrature the eight systematic as well as the statistical uncertainties quoted in the files {\tt Pk1D\_syst.dat} and {\tt Pk1D\_data.dat}, respectively. Furthermore, we performed the fits (i) assuming diagonal covariance and (ii) using the covariance matrices reported in {\tt Pk1D\_cor.dat}, respectively. We find that (i) yields slightly more conservative results and therefore quote this case for our fiducial result; we refer to the discussion at the end of section~\ref{sec:robust} for details.},
covering the range $0.001-0.02($km/s$)^{-1}\sim 0.1-2\,h/$Mpc and $2.2\leq z\leq 4.6$. As mentioned before, we restrict our analysis to the redshift
bins $z=3.0, 3.2, 3.4, 3.6, 3.8, 4.0, 4.2$, omitting lower redshifts (for which non-linearities are more pronounced) and the highest bins (due
to an increased sensitivity to reionization physics). Furthermore, we consider the baseline  model for the 1D flux power spectrum described by the six parameters \eqref{eq:baseline}
that are all left free in the fit, as well as the restricted model for which $A$ is given by \eqref{eq:A} with $c=0.8$ determined by a calibration with simulation data. 
At the end of this section, we discuss the dependence on various assumptions and their impact on the neutrino mass bound.
The main purpose of this work is to demonstrate that the analytical Lyman-$\alpha$ model can be used to obtain conservative bounds on the sum of neutrino masses,
while marginalizing over IGM parameters. Therefore, as for the comparison to simulation data, we have fixed all cosmological parameters, except for
the neutrino mass, as given in the beginning of section \ref{sec:val}. This restriction should be kept in mind when interpreting the numerical value of the
neutrino mass bounds quoted below. 

\subsection{Fit of the effective model to BOSS data} 

\begin{figure}[t!]
        \centering
        \includegraphics[width=0.65\textwidth]{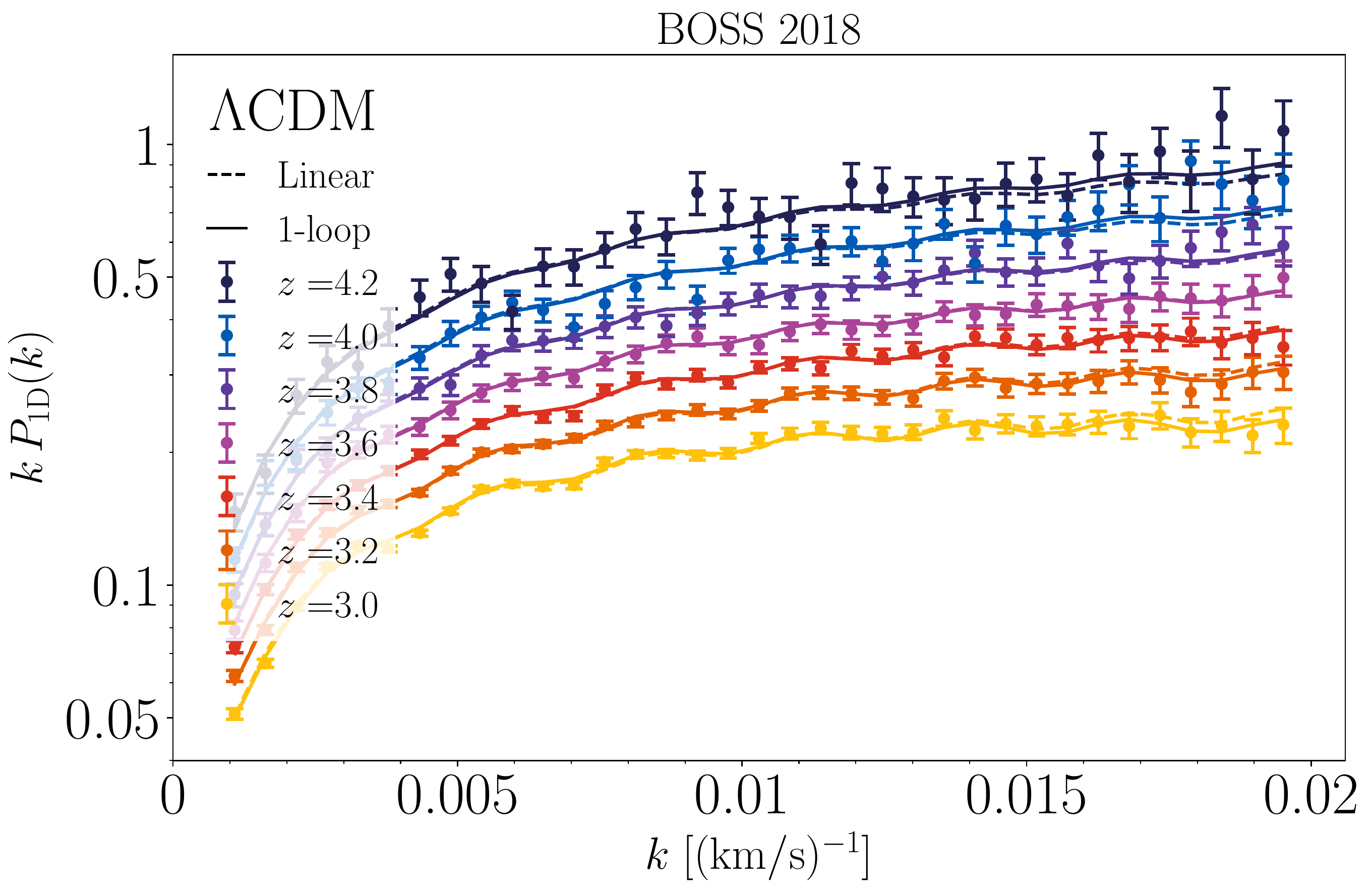}
        \caption{1D Lyman-$\alpha$ flux power spectrum from BOSS \cite{Chabanier:2018rga} (data points) and best-fit $\Lambda$CDM analytical model (lines). Solid lines correspond to 1-loop input power spectra, and dashed lines to the linear approximation. 
}
        \label{fig:pk_boss}
\end{figure}

In figure~\ref{fig:pk_boss}, we show the best-fit $\Lambda$CDM model together with the BOSS data~\cite{Chabanier:2018rga}. We observe that the analytical model with 1-loop input power spectra yields a valid description of the 1D flux power spectrum, with a total $\chi^2=193.4$. This can be compared to the number of degrees of freedom, given by $35$ $k$-bins $\times 7$ redshifts, and subtracting six free model parameters, giving $239$. As expected, the total $\chi^2$ value is significantly larger as for the
simulation data. 
For comparison, also the result when using linear instead of 1-loop input spectra is shown with dashed lines in figure~\ref{fig:pk_boss}. We find a larger value
($\chi^2=206.5$) when using linear instead of 1-loop input power spectra. 
Thus, including non-linear corrections in the input power spectrum improves the fit, similarly as observed for the simulation data.
\begin{figure}[t]
        \centering
        \includegraphics[width=0.65\textwidth]{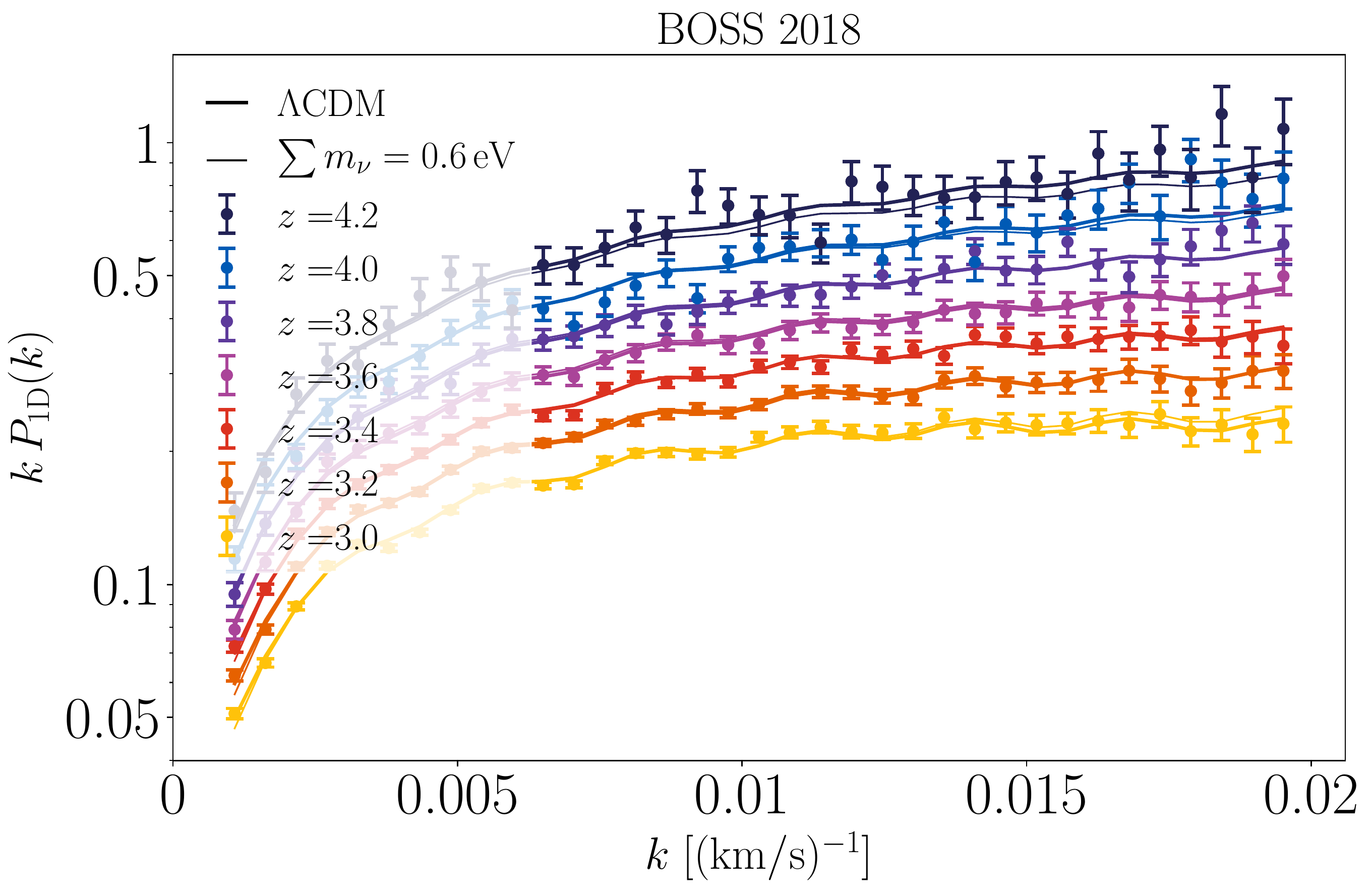}
        \caption{1D Lyman-$\alpha$ flux power spectrum from BOSS \cite{Chabanier:2018rga} (data points) and best-fit analytical models when using 1-loop power spectra as input (lines) and fixing the amplitude. Thick and thin lines show the cases $\sum m_\nu=0$\,eV ($\Lambda$CDM) and $\sum m_\nu=0.6$\,eV, respectively.
}
        \label{fig:pk_boss_LCDM_vs_mnu0p6}
\end{figure}

In figure~\ref{fig:pk_boss_LCDM_vs_mnu0p6}, we show the best-fit analytical model for the cases of $\sum m_\nu=0,0.6$\,eV when using a 1-loop input power spectrum and a fixed amplitude. From there we can already see that for large wavenumbers the $\Lambda$CDM model yields a better fit to the BOSS data than the one with $\sum m_\nu=0.6$\,eV. Indeed, we find a total value of $\chi^2=193.4$ in the former and $\chi^2=230.2$ in the latter case. (Note that for $\Lambda$CDM the case with fixed and free amplitude coincide).
In contrast to this, when using linear input power spectra, we obtain a value $\chi^2=207.8$ for $\sum m_\nu=0.6\,$eV that is very similar to the one for $\Lambda$CDM, $\chi^2=206.5$.
Accordingly, the sensitivity to the neutrino mass is much higher at 1-loop order than in the linear approximation.

\begin{figure}[tb]
        \centering
        \includegraphics[width=0.65\textwidth]{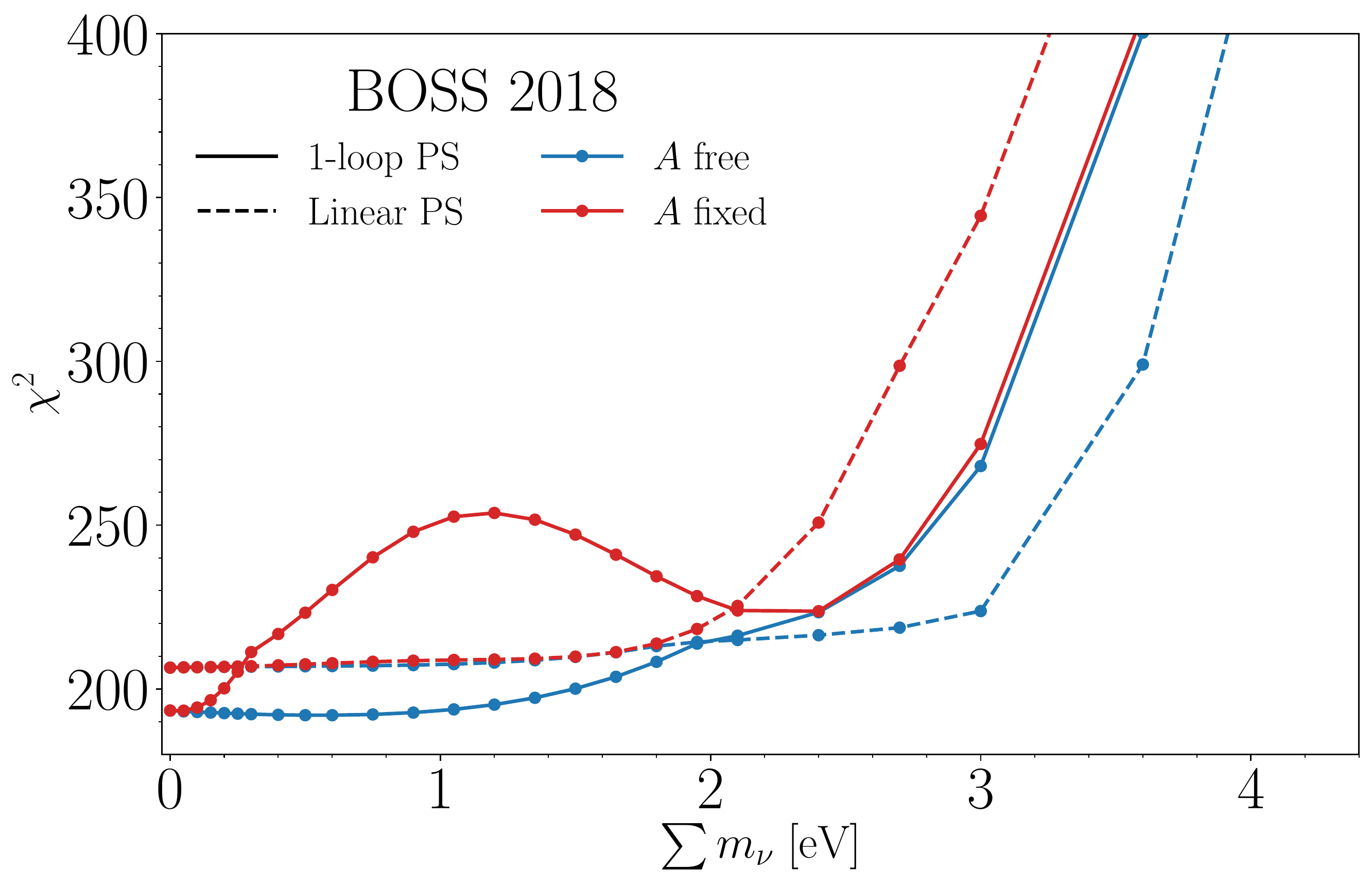}
        \caption{
$\chi^2$ obtained from a fit of the analytical model to BOSS data~\cite{Chabanier:2018rga}. Blue lines correspond to the baseline analytical model with free amplitude, while $A$ is fixed according to \eqref{eq:A} for the red lines. Solid lines correspond to 1-loop, and dashed to linear input power spectra.
}
        \label{fig:chi2_boss}
\end{figure}

The dependence of $\chi^2$ on the neutrino mass is shown in figure~\ref{fig:chi2_boss}. We show the result obtained when using 1-loop or linear input power spectra, as well
as for the baseline model and the case with fixed amplitude $A$, respectively. As expected, including the 1-loop correction is crucial for the sensitivity to the neutrino mass.
In addition, as discussed above, fixing $A$ by calibrating with simulation data breaks the degeneracy between the neutrino mass and the amplitude. The shape of the $\chi^2$ curves is similar as for
the fit to simulation data, apart from the overall offset in the total value of $\chi^2$.
Note that the results obtained with linear input power spectra in figure~\ref{fig:chi2_boss} are included for illustrative purposes only, and
we focus on the 1-loop case in the following. In addition, we use $\Delta\chi^2$ for a relative comparison of different models.

\subsection{Neutrino mass bound} 

The BOSS data are compatible with massless neutrinos.
To extract a $95\%$ C.L. upper bound on the neutrino mass, we require
that $\Delta\chi^2<3.84$ compared to the minimal value. 
Taking the case with 1-loop input power spectra as well as fixed amplitude $A$ as our fiducial choice, we extract a nominal $95\%$ C.L. upper limit of:
\be
  \sum m_\nu \leq 0.16\,{\rm eV}.
\ee
Even though we keep the cosmological parameters fixed in this study, it is instructive to compare the upper bound to those derived in~\cite{Palanque-Delabrouille:2019iyz}
based on the full set of BOSS data~\cite{Chabanier:2018rga}, as well as a suite of hydrodynamical simulations in order to predict the flux power spectrum. 
When combining Lyman-$\alpha$ with CMB temperature and polarisation data from Planck~\cite{Aghanim:2018eyx}, the  $95\%$ C.L. bound lies in the range $0.10-0.13$\,eV.
While \cite{Palanque-Delabrouille:2019iyz} finds a slight tension between Planck and BOSS Lyman-$\alpha$ data (that can be improved when including a running spectral index
in the cosmological model), there is overall a good agreement. In our analysis, the Planck results enter indirectly via the fixed set of cosmological parameters, in particular the
normalization of the primordial power spectrum $A_s$. 
We leave an analysis with a combined fit of IGM and cosmological parameters to future work. Given that the analytical model allows for a considerable freedom regarding the impact of
the IGM, the upper bound can be considered as conservative. Nevertheless, some input from simulations is required to calibrate the parameter $c$
entering the relation \eqref{eq:A} for the amplitude $A$. Further work is required to determine the sensitivity of this relation to changes in the cosmological parameters.

When using Lyman-$\alpha$ data only, together with a prior $H_0=67.3\pm 1.0\,{\rm km}/{\rm s}/{\rm Mpc}$, the upper bound at $95\%$ C.L. is found to be $\simeq 0.58-0.71$\,eV 
in~\cite{Palanque-Delabrouille:2019iyz}. The main reasons for the large improvement when combining with CMB data is that the approximate degeneracy between $A_s$ and $\sum m_\nu$ is broken.
Within the analytical model considered here, a similar degeneracy between the amplitude parameter $A$ and the neutrino mass occurs.  Accordingly, when leaving the amplitude $A$ free, the bound weakens significantly, to $\sum m_\nu \leq 1.24\,{\rm eV}$.
 Note that the Lyman-$\alpha$ model parameter $A$ describes the overall normalization of the non-linear 1D flux power spectrum, while $A_s$ corresponds to the usual $\Lambda$CDM parameter related
to the normalization of the linear 3D matter power spectrum. Therefore, they are distinct parameters, and, depending on the properties of the IGM, $A$ can vary even when $A_s$ is fixed.
Nevertheless, the scenario where $A$ is left completely free, while fixing $A_s$, should be considered as extremely conservative.
The upper bound obtained for a free amplitude is consistent with the expectation from the simulation data studied in the previous section, and related to the fact that for very large values of the neutrino mass, the neutrino power spectrum becomes strongly scale-dependent within the $k$-range measured by BOSS.

\begin{figure}[t]
        \centering
        \includegraphics[width=\textwidth]{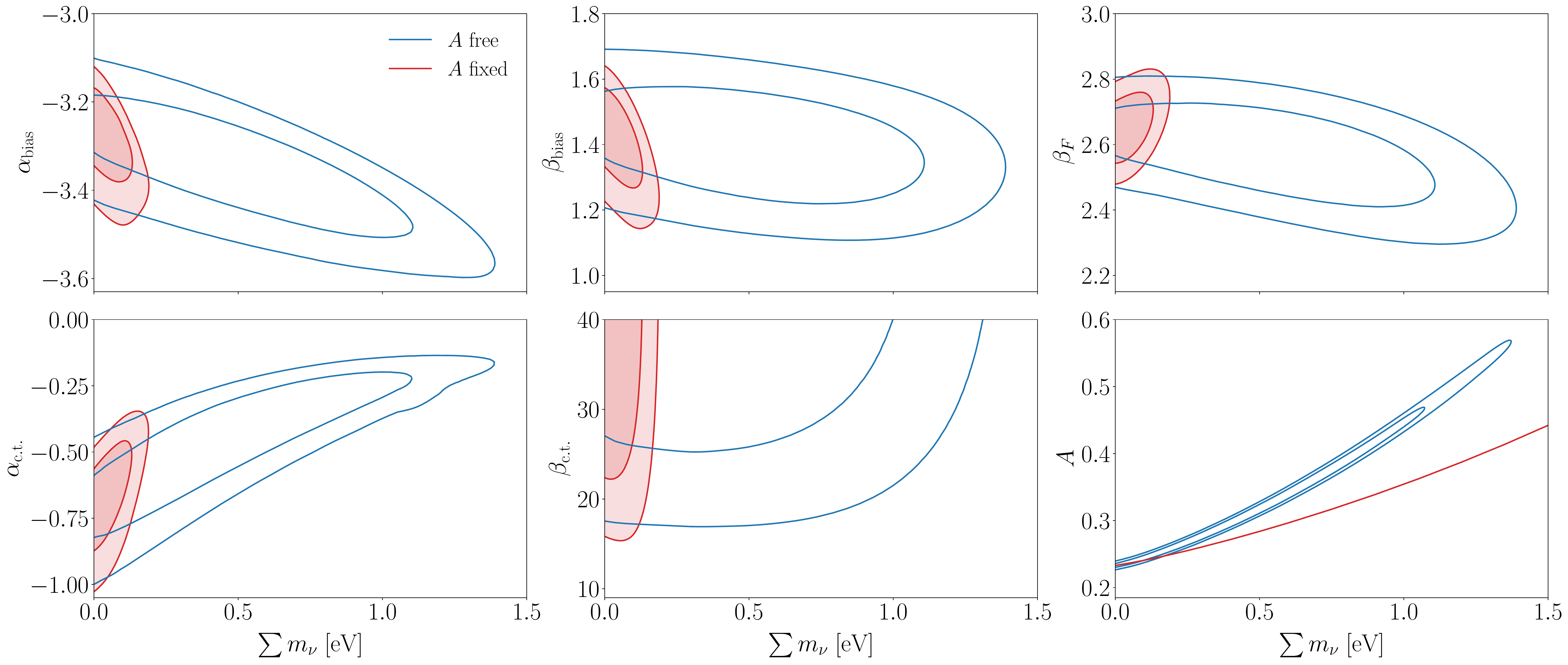}
        \caption{Two-dimensional $95\%$ and $68\%$ C.L. contours for each of the effective model parameters that are left free in the fit, and the sum of neutrino masses. Blue lines correspond to the baseline model with free amplitude, and red-shaded contours to the case with amplitude fixed by calibration with simulations according to \eqref{eq:A}. For the lower right panel, only blue contours are shown, since $A$ is not a free model parameter for the case of fixed amplitude. Instead, the red line in the lower right panel shows the dependence of $A$ on $\sum m_\nu$ implied by \eqref{eq:A}.}
        \label{fig:2dgrid}
\end{figure}

Once the degeneracy between the sum of neutrino masses and the overall amplitude $A$ is lifted by \eqref{eq:A}, the remaining free parameters are well constrained, and no further significant degeneracy with the neutrino mass remains. This can be seen in figure~\ref{fig:2dgrid}, where two-dimensional confidence contours for the case with fixed amplitude are shown in red. For comparison, also the
(much larger) regions obtained when letting the amplitude free are shown by the blue lines. The degeneracy between $A$ and $\sum m_\nu$ can in particular be seen in the lower right panel. In addition, the red line in the lower right panel shows the relation between the amplitude $A$ and the sum of neutrino masses \eqref{eq:A} obtained from the calibration with simulations. The increase in sensitivity for the model with fixed amplitude is related to the different slope of this line as compared to the narrow blue confidence regions, along which the neutrino mass is degenerate with $A$. We also note that the parameter $\beta_{\rm c.t.}$, which determines the redshift-dependence of the counterterm (cf. \eqref{eq:alphabeta}), is compatible with large values. This implies that the counterterm-contribution to $I_0$ is mostly relevant towards the lowest redshifts considered in the fit, while it is strongly suppressed at high redshifts. The two-dimensional confidence contours are obtained by minimizing $\chi^2$ over the remaining 4 (5) free parameters for fixed (free) amplitude, and requiring $\Delta\chi^2 < 2.28$ or $5.99$ relative to the global best fit at $68\%$ or $95\%$ C.L., respectively, as appropriate for a $\chi_2^2$ distribution. The latter occurs for $\sum m_\nu=0.028 (0.55)$\,eV with $\chi^2= 193.28 (191.96)$ for fixed (free) amplitude, but is well compatible with both massless neutrinos (i.e. $\Lambda$CDM, $\chi^2=193.39 (193.39)$) as well as $0.05$\,eV ($\chi^2= 193.36 (193.13$)) at $1\sigma$.

\begin{figure}[t]
        \centering
        \includegraphics[width=0.65\textwidth]{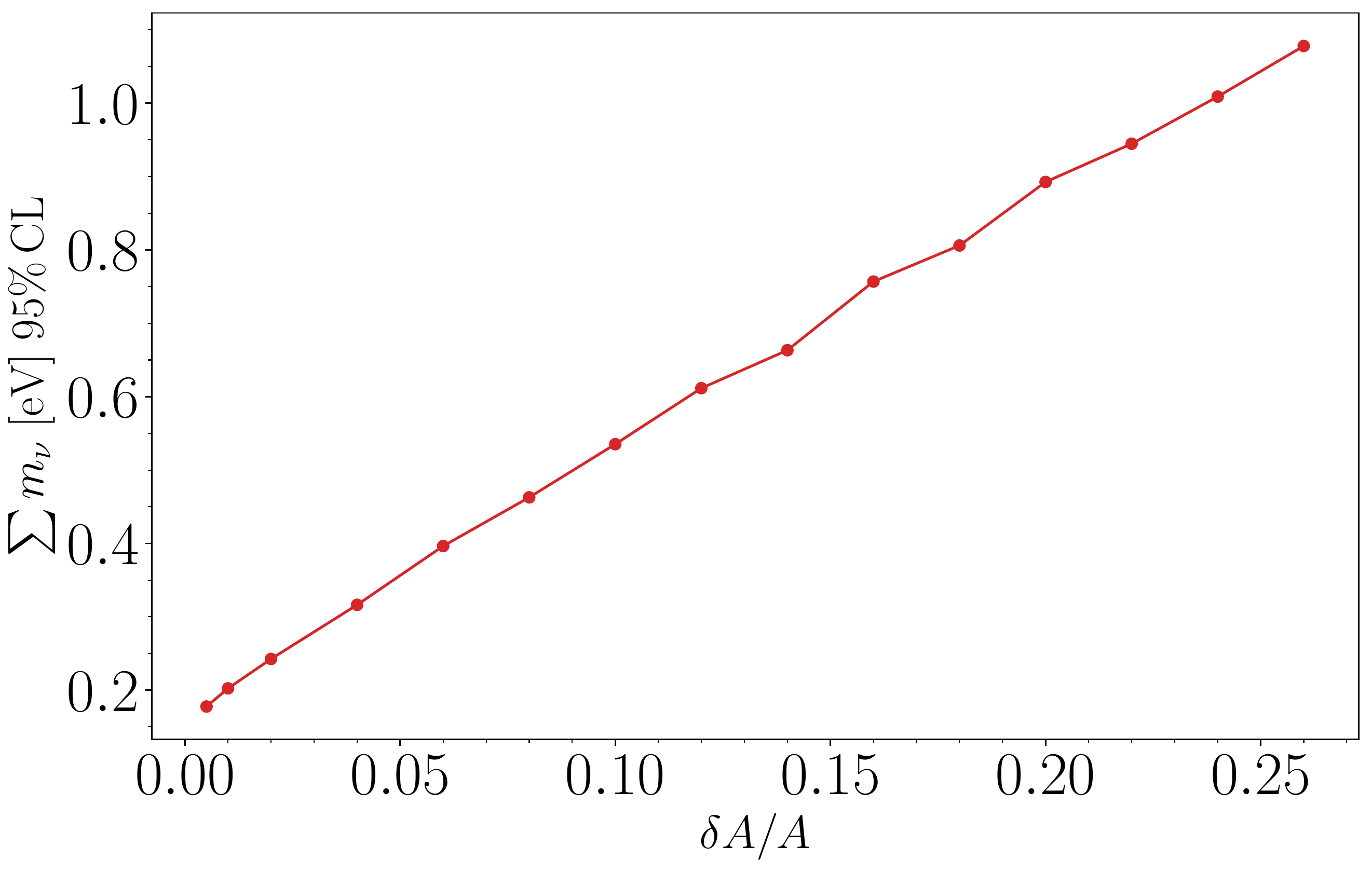}
        \caption{$95\%$ C.L. upper bound on $\sum m_\nu$ when allowing for a tolerance in the amplitude, $|A-A_{\rm fix}|\leq \delta A$, where $A_{\rm fix}$ is given by \eqref{eq:A}. The limit $\delta A/A\to 0$
corresponds to the baseline model with fixed amplitude, and the opposite limit to the case where $A$ is left completely free.}
        \label{fig:dAA}
\end{figure}

\subsection{Robustness and dependence on assumptions}\label{sec:robust} 

We now turn to the discussion of the impact of various assumptions on the neutrino mass bound. 
The most relevant is the relation \eqref{eq:A} for the dependence of the overall amplitude $A$ on the neutrino mass, that was calibrated from hydrodynamical simulation data. 
So far we have either assumed that $A$ is completely free, or entirely fixed according to \eqref{eq:A}.
In order to quantify by how much the neutrino mass bound relaxes when allowing for some freedom in the amplitude $A$, we have considered an intermediate scenario, where we
allow for a relative variation of $\delta A/A$ above or below the fiducial value~\eqref{eq:A}. The dependence of the upper bound on $\delta A/A$ is shown
in figure~\ref{fig:dAA}. For $\delta A/A\to 0$ one recovers the case with fixed amplitude, while for the largest value $\delta A/A=26\%$ shown in figure~\ref{fig:dAA}, the upper bound is
already close to that with completely free amplitude. For $\delta A/A=5\%$ the upper bound degrades from $0.16\,$eV to $0.35\,$eV. Therefore, an accurate control over the overall
amplitude of the 1D flux power spectrum (relative to the total flux) is crucial for the robustness of the neutrino mass bound. 
We checked that, when using $\Lambda$CDM simulation results instead of BOSS data, we obtain a dependence
comparable to the one in figure~\ref{fig:dAA}.

In order to estimate uncertainties on $A$ due to our incomplete knowledge of the thermal history of the IGM, we performed a test using a set of hydrodyamical simulations that all correspond to the same set of cosmological parameters (identical to the ones assumed in section~\ref{sec:val}, and for massless neutrinos). However, the thermal histories of the IGM are different. We adopt a ``cold'' and a ``hot'' scenario for the IGM temperature, that corresponds to $T=1.1\, (2.3) \cdot 10^4$\,K at $z=3$, compared to the reference case with $T=1.6\cdot 10^4$\,K (these values refer to the temperature at the mean IGM density). These models {\it conservatively} bracket the observed temperature ranges for the IGM. We then fit the fiducial effective model to each of the simulations, within the same $k$ and $z$ range as before. We find that the 1D flux power spectra can be well described by the effective model in each case, i.e.~the dependence on the thermal history can indeed be absorbed into shifts of the free model parameters. In particular, the counterterm and velocity bias parameters change by ${\cal O}(1)$, which is expected since the IGM temperature varies by more than a factor of two for the cold and hot scenarios. In contrast, the best-fit value for $A$ changes at the level of less than $10\%$ as compared to the reference case. This means that uncertainties related to the IGM evolution are mostly absorbed by the counterterm and velocity bias contributions. Nevertheless, an uncertainty in $A$ of order $10\%$ potentially compromises the ability to constrain the neutrino mass, when following a conservative approach that allows for marginalization over variations in the IGM temperature bracketed by the hot and cold scenarios. Therefore, a detailed modelling of the thermal evolution will be important, especially if smaller scales need to be addressed. We postpone this to a future publication in which we  will also discuss the comparison with future Lyman-$\alpha$ observations, which will be helpful to mitigate the impact of IGM uncertainties on the neutrino mass bound (also by relying on astrophysical priors). 

In addition, we performed an analogous check for a set of hydrodynamical simulations in which the temperature-density relation is varied (corresponding to values of the adiabatic index of $\gamma=1.0 (1.6)$, respectively, instead of $1.4$ for the reference case). In this case we find a smaller impact on $A$, of the order of $2.5\%$. Lastly, we considered a variation in the redshift of reionization, $z_r=5.4 (7.4)$, while $z_r=6.0$ in the reference model. The impact on $A$ is almost negligible, at the $0.2\%$ level.

\begin{figure}[t]
        \centering
        \includegraphics[width=0.65\textwidth]{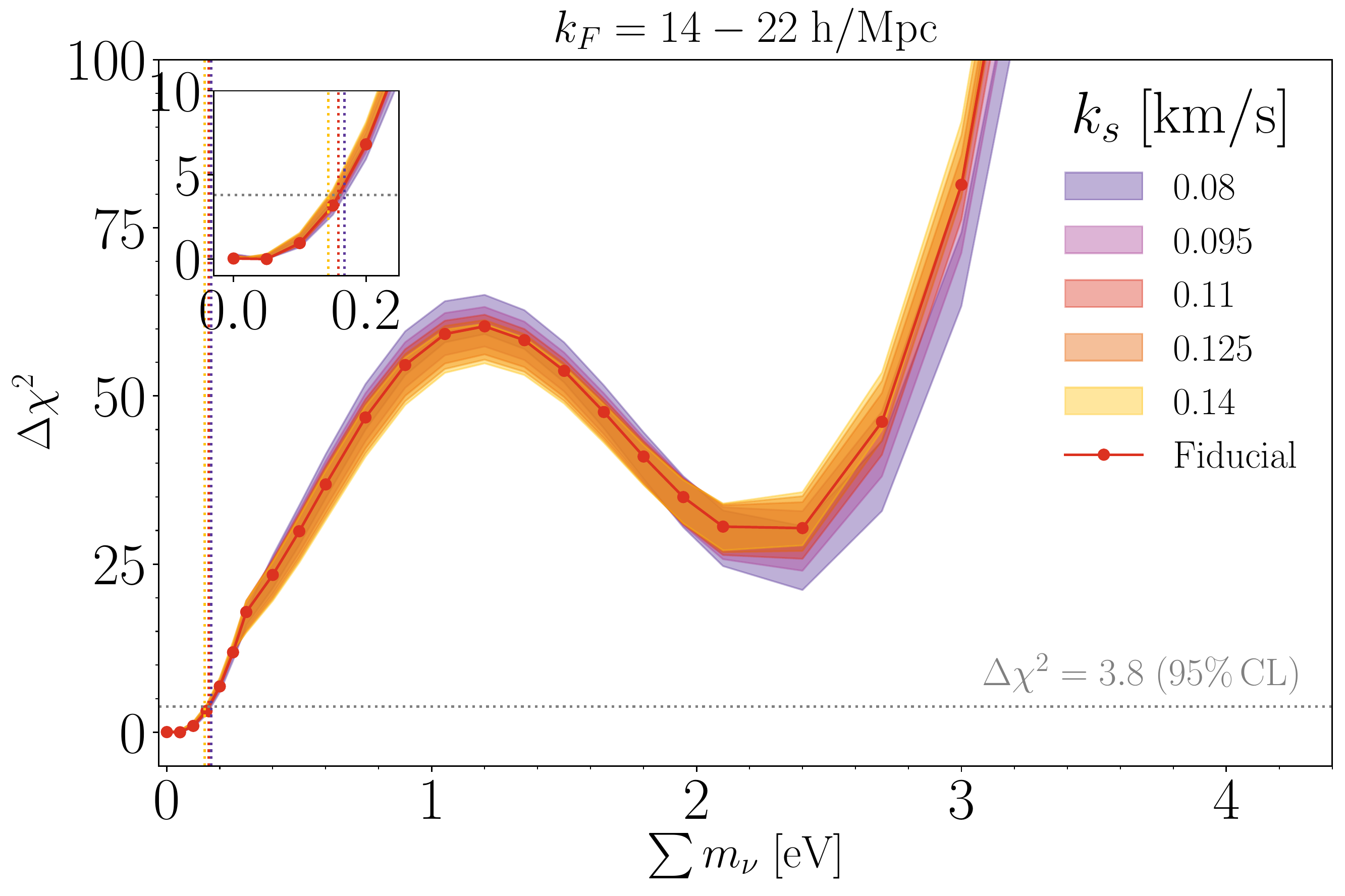}
        \caption{Dependence of $\Delta \chi^2$ on the parameters $k_s$ and $k_F$ obtained from a fit of the analytical model with fixed amplitude and 1-loop input spectrum to BOSS data~\cite{Chabanier:2018rga}. The shaded regions show the maximal and minimal  $\Delta \chi^2$ values when varying $k_F=14-22 h/$Mpc, for given values of $k_s$ (see legend). The $\Delta \chi^2$ for the baseline model with $k_F=\unit[0.11]{km/s}$ and $k_F=\unit[18]{h/Mpc}$ is also shown for comparison. The vertical lines indicate the lowest and highest value of the $95\%$ C.L. bound on $\sum m_\nu$ obtained for any set $(k_F,k_s)$ within the ranges indicated in the figure, respectively, as well as the bound obtained for the baseline model.}
        \label{fig:kskF}
\end{figure}

Next, we quantify in how far the parameters $k_s$ and $k_F$, related to thermal broadening as well as the Jeans scale due to baryonic pressure, respectively, influence the
neutrino mass bound. We stress that, within the analytical model considered here, the impact of the unknown IGM physics is mainly accounted for by the bias and counterterm
parameters, that are marginalized over in all cases. The impact of the choice of $k_s$ and $k_F$ on the 1D flux power spectrum on BOSS scales is only minor.
For our fiducial analysis, these parameters were therefore fixed to $k_F=18h/$Mpc and $k_s=0.11({\rm km}/{\rm s})^{-1}$, respectively. In figure~\ref{fig:kskF}, we show the envelope
of the $\chi^2$ curves obtained when varying $k_F$ in the range $14-22h/$Mpc, for various values of $k_s$ within $0.08-0.14({\rm km}/{\rm s})^{-1}$. The corresponding neutrino mass bound,
extracted for a grid of fixed values of $k_s$ and $k_F$, chosen within the ranges given above, is always close to the fiducial value $0.16\,$eV, with the smallest and largest values being $0.144$ and $0.167$\,eV, respectively.
Alternatively, when marginalizing over $k_s$ and $k_F$ within the same ranges (i.e. treating these parameters as free values in the fit, and minimizing $\chi^2$), the resulting mass
bound is found to be $0.15\,$eV. We conclude that the sensitivity to the model parameters $k_s$ and $k_F$ is minor.

As for the simulation data, we also checked that our results are robust when varying the cutoff that is
imposed in the computation of the integrals $I_{0,2,4}$  entering the 1D power spectrum, see~\eqref{eq:I024}.
When varying the cutoff in the range $10-20 h/$Mpc, the neutrino mass bound changes within $0.153-0.159$\,eV.
This check indicates that the counterterm and bias parameters are indeed suitable to absorb the unknown UV contributions 
to the 1D flux power spectrum. We also verified that, within the scope of our analysis, we obtain stable results when computing $\chi^2$ by summing over all $k$-bins
individually, or taking the full covariance matrices as provided by BOSS~\cite{Chabanier:2018rga} into account. Since the former leads to slightly more conservative
results, we adopted this choice for our fiducial analysis. Otherwise, the neutrino mass bounds slightly improve to $0.14$\,eV and $1.10$\,eV for fixed or free amplitude $A$, respectively.

Finally, we checked that when using the cold dark matter and baryon (cb) power spectra as input for the effective model on \emph{both} the linear and one-loop level (see~\eqref{eq:P1loop}), as suggested by the analysis of the halo power spectrum in~\cite{Villaescusa-Navarro:2017mfx}, only minor differences occur. In particular, the neutrino mass bound obtained in this case with fixed amplitude is practically unaffected (being $0.159$\,eV in both cases) while the upper bound with free amplitude shifts from $1.24$\,eV to $1.40$\,eV. Further investigation of this point in the future would be interesting.

\subsection{Prospects} 

Future surveys such as DESI \cite{Aghamousa:2016zmz} will observe quasar absorption spectra and provide precise measurements of the 1D Lyman-$\alpha$ forest flux power spectrum~\cite{Karacayli:2020aad}. 
In order to provide a rough estimate for potential future improvements of the neutrino mass bound based on the effective model approach, we consider a setup with the same redshift and $k$-range as applied to the BOSS analysis, but reduced statistical and/or systematic uncertainties. In particular, when assuming the same systematic errors as for the BOSS data~\cite{Chabanier:2018rga}, but a statistical error that is smaller by a factor of two, we find an improvement in the $95\%$C.L. neutrino mass bound from $0.16$ to $0.14$\,eV. When assuming that both the statistical and systematic errors can be reduced by a factor two, the
projected bound further decreases to $0.10$\,eV. 
This limit is promising since it is likely that the improvement over systematic and statistical errors will be soon achieved, with the DESI data set. It is also expected that this limit can be improved further by complementing the 1D flux power with: 3D flux power information; higher order statistics like the bispectrum; other external intergalactic medium data sets like high and medium resolution quasar spectra that will allow to break the internal degeneracies with better measurements of astrophysical and nuisance parameters of the models; external cosmological data sets like Baryonic Acoustic Oscillations measurements.

As a somewhat more aggressive alternative, we follow~\cite{Walther:2020hxc} and use an estimate of $1\%$ for the relative error at all $k$- and $z$-values. We keep the same range for $k$ and $z$ as in our previous analysis, covering the region of validity of the effective model. Based on the simulation data, we find an expected $95\%$C.L. upper limit of $0.056$\,eV provided the true value of the sum of neutrino masses would be zero. For a ``true'' value of $0.15$\,eV, the neutrino mass can be determined at $95\%$C.L. with a relative uncertainty of $17\%$. We checked that the relative uncertainty is around $17-23\%$ for all simulated neutrino masses that we considered (being $0.15, 0.3,0.6,0.9$\,eV).

\section{Summary \label{sec:sum}} 

We presented an effective setup to model the flux power spectrum of the Lyman-$\alpha$ forest. The model is inspired by early 
analytical models using the Zel'dovich approximation but encodes the complicated dynamics of the IGM into a few effective parameters. One novel
ingredient in our model is that the UV dependence of the flux power spectrum is absorbed into a counter term. Overall, this model has only six relevant parameters
that are sufficient to fit to simulation or BOSS data to very high precision (within the observed range, far above the Jeans scale). 

The main goal of the present work is to assess the predictive power of the model in the context of neutrino masses. 
When confronted with real data, the model displays a degeneracy between the normalization of the flux power spectrum and the sum of the neutrino masses. 
A similar degeneracy is also found in the Lyman-$\alpha$ analysis using simulation data. 
We removed this ambiguity by calibrating one of the model parameters to simulations. After the degeneracy is lifted, the model predicts an upper bound on the 
neutrino masses, $\sum m_\nu \leq 0.16\,{\rm eV}$ ($95\%$ C.L.), when confronted with BOSS data~\cite{Chabanier:2018rga} for $3\leq z\leq 4.2$. 
Notice that this result is based on the 1-loop power spectrum as input to our model, while the 
linear power spectrum would not allow to deduce any stringent bound on the neutrino masses. In addition, while marginalizing over the model parameters that capture the
unknown IGM dynamics, we kept the cosmological parameters fixed (in agreement with Planck CMB data) in this study, and applied a simplified treatment of the statistical and systematic uncertainties of BOSS data. 
Nevertheless, our results can be regarded as a proof-of-principle for obtaining conservative constraints from Lyman-$\alpha$ forest observations based on effective theory methods
and semi-analytical models.
In particular, the low computational cost of the effective model is suitable for applying parameter estimation methods based on Monte Carlo sampling. For a single set of
cosmological input parameters, the computational complexity is comparable to a standard 1-loop computation in perturbation theory. Within our implementation, the
time required to produce the relevant power spectra for all $k$ and $z$ values is of the order of minutes on a standard desktop pc. We expect that this can be further reduced by applying
fast Fourier techniques analogously to those described in \cite{Schoneberg:2018fis,Chudaykin:2020aoj} and used in Monte Carlo analyses of BOSS galaxy clustering data
based on effective theory \cite{Ivanov:2019hqk}. For comparison, a Monte Carlo analysis of Lyman-$\alpha$ data based on a grid of hydrodynamical simulations
can take of the order of a month of CPU time.

In principle, the same strategy can also be followed to obtain an effective model for the 3D spectrum of the optical depth. 
However, this requires to introduce a large number of additional counterterms in the analysis, see~\cite{Desjacques:2018pfv,Cabass:2018hum}, that are
degenerate with the counterterms considered here as far as the 1D flux power spectrum is concerned. While such an extended model would loose predictive power
when using observational data for the 1D spectrum only, it would be interesting to investigate whether this approach can be used
to simultaneously describe the 3D and 1D flux power spectra on large scales.

\section*{Acknowledgments}

We thank Simeon Bird, Vid Ir\v si\v c, Julien Lesgourgues, Andreu Font-Ribera and Nils Sch\"oneberg for helpful discussions. This work is supported by the Deutsche Forschungsgemeinschaft (DFG, German Research Foundation)
under Germany's Excellence Strategy -- EXC 2121 ``Quantum Universe'' -- 390833306 as well as -- EXC-2094 ``Universe'' -- 390783311, through
the grant CRC-TR 211 ``Strong-interaction matter under extreme conditions'', the Emmy Noether grant No. KA 4662/1-1, and
by the Munich Institute for Astro- and Particle Physics (MIAPP) which is funded by the DFG -- EXC-2094 -- 390783311.
MV is supported by INFN INDARK grant and  by a grant from the agreement ASI-INAF n.2017-14-H.0.
Part of the simulations used in this project were run on the Ulysses supercomputer at SISSA. 
The Sherwood and SherwoodRelics simulations that were used in this work were performed with supercomputer time awarded by the Partnership for Advanced Computing in Europe (PRACE) 8th and 16th calls. We acknowledge PRACE for awarding us access to the Curie and Irene supercomputers, based in France at the Tr´es Grand Centre de Calcul (TGCC).


\end{document}